\documentclass[%
 aip,
jmp,
 amsmath,amssymb,
 reprint,%
]{revtex4-1}

\usepackage{graphicx}
\usepackage{dcolumn}
\usepackage{bm}
\usepackage[mathlines]{lineno}
\usepackage[colorlinks=true,linkcolor=green,citecolor=blue]{hyperref} 
\usepackage[utf8]{inputenc}
\usepackage[T1]{fontenc}
\usepackage{mathptmx}
\usepackage{etoolbox}
\usepackage{soul}

\makeatletter
\def\@email#1#2{%
 \endgroup
 \patchcmd{\titleblock@produce}
  {\frontmatter@RRAPformat}
  {\frontmatter@RRAPformat{\produce@RRAP{*#1\href{mailto:#2}{#2}}}\frontmatter@RRAPformat}
  {}{}
}%
\makeatother
\begin{document}


\title[Magnetic reconnection in the cool low solar atmosphere]{High $\beta$ magnetic reconnection at different altitudes in the cool low solar atmosphere}
\author{Abdullah Zafar}
\affiliation{Yunnan Observatories, Chinese Academy of Sciences, Kunming, Yunnan 6502016, PR China}%

\author{Lei Ni}
\email{leini@ynao.ac.cn}
\affiliation{Yunnan Observatories, Chinese Academy of Sciences, Kunming, Yunnan 6502016, PR China}
\affiliation{Center for Astronomical Mega-Science, Chinese Academy of Sciences, 20A Datun Road, Chaoyang District, Beijing 100012, PR China}
\affiliation{University of Chinese Academy of Sciences, Beijing 100049, PR China}

\author{Jun Lin}
\affiliation{Yunnan Observatories, Chinese Academy of Sciences, Kunming, Yunnan 6502016, PR China}
\affiliation{Center for Astronomical Mega-Science, Chinese Academy of Sciences, 20A Datun Road, Chaoyang District, Beijing 100012, PR China}
\affiliation{University of Chinese Academy of Sciences, Beijing 100049, PR China}

\author{Udo Ziegler}
\affiliation{Astrophysikalisches Institut Potsdam D-14482 Potsdam, Germany}

\date{\today}

\begin{abstract}
We numerically studied magnetic reconnection in a high $\beta$ hydrogen-helium plasma at different altitudes from the photosphere to the upper chromosphere. The time dependent ionization degrees were included to get more realistic diffusivities and viscosity, and
appropriate radiative cooling models were applied. 
Our numerical results indicate that the plasmoid instability always plays a vital role in speeding up magnetic reconnection at different atmospheric layers. 
In addition, both the strong radiative cooling and the magnetic diffusion caused by the electron-neutral collision ($\eta_{en}$) can significantly accelerate magnetic reconnection below the middle chromosphere. 
On the other hand, both the ambipolar diffusion and the viscosity result in higher temperature and plasma pressure in the reconnection region in the upper chromosphere, which then hinder the fast reconnection process from developing. 
The local compression heating triggered by turbulent reconnection mediated with plasmoids is the dominant heating mechanism in the unstable reconnection stage at different atmospheric layers, but the viscous heating and the ambipolar diffusion heating are equally important in the upper chromosphere. 
The Joule heating contributed by $\eta_{en}$ dominates during the early quasi-steady reconnection stage below the middle chromosphere, the strong radiative cooling also leads to much stronger compression heating and more generation of thermal energy in this region. 
Though the plasma $\beta$ is the same in all the simulation cases at different altitudes, the temperature increase is more significant in the upper chromosphere with much lower density and weaker radiative cooling.

\end{abstract}

\maketitle
\section{Introduction}
Magnetic reconnection is a fundamental plasma phenomena in which antiparallel magnetic field lines come closer together, break, and then rejoin, resulting in a modified topological configuration. During this process, the magnetic energy is converted into kinetic energy and thermal energy of the plasma. 
Magnetic reconnection has been observed in different environments, such as, astrophysical plasmas~\cite{shibata2011solar,su2013imaging}, the Earth’s magnetotail~\cite{retino2007situ}, and magnetically confined plasmas in fusion devices~\cite{hastie1997sawtooth,yamada2011mechanisms}. 
The transient and energetic events observed in the solar atmosphere, ranging from flares~\cite{lin2005direct} and coronal mass ejections~\cite{lin2005direct,lin2007features} on the larger scales to  jets~\cite{shen2021observation,sterling2016minifilament,shen2017solar} and microflares~\cite{brosius2009observations,kirichenko2017plasma,wright2017microflare} on spatial sizes of only a few arcseconds , are believed to be driven by magnetic reconnection.

Such wide range of reconnection events ought to be fast in order to serve as an energy release mechanism. 
However, observed spatial and temporal scales and plasma conditions in which they take place indicates that the Sweet-Parker model~\cite{parker1957sweet,1958IAUS....6..123S} acts too slowly to be responsible for the rapid energy release. 
The Petschek model achieves fast reconnection rates. But the local enhancement of the resistivity is required to keep the Petschek type reconnection. 
The Petschek length in astrophysical setting is usually smaller than the mean free path of the electron-ion collision, which implies that the collisionless effects could be important before reaching the Petschek scale~\cite{zweibel2009magnetic}. 
Further more, these classical steady-state reconnection models usually do not match the widely observed impulsive reconnection processes in astrophysical environments. 
Previous theoretical studies revealed that Sweet-Parker reconnection may evolve to fast magnetic reconnection due to the fragmentation of the current sheet and the development of secondary islands, the plasmoid instability~\cite{loureiro2007instability,bhattacharjee2009fast}. 
The plasmoid instability occurs as the Lundquist number exceeds a critical value of around 10$^{4}$ ~\cite{biskamp1986magnetic,samtaney2009formation,bhattacharjee2009fast,ni2010linear}, which is way below the Lundquist number in the solar atmosphere. 
Bright blob-like structures (plasmoids) have been found in various dynamical events in the solar atmosphere~\cite{ko2003dynamical,zhang2016observations,singh2012multiple,li2016magnetic}. 
Turbulent magnetic reconnection mediated by plasmoid instability in the solar corona has been frequently studied recently \cite{ni2012impact,ye2019numerical,mei2017magnetic,wang2022current,zhao2021magnetic,dong2022reconnection,zhu2020relativistic,zhu2020relativistic2}.

The lower solar atmosphere is complex, inhomogeneous and highly dynamic. 
It contains plasma with a wide range of density, temperature, and ionization. 
The hydrogen density drops with height by several order of magnitudes from the photosphere to the bottom of the corona. The ionization degree in the lower solar atmosphere is small, down to $10^{-4}$ in the temperature minimum region (TMR).
The small scale reconnection event such as chromospheric jets~\cite{shibata2007chromospheric,singh2011chromospheric}, Ellerman bombs (EBs)~\cite{ellerman1917solar,georgoulis2002statistics,reid2016magnetic}, Ultraviolet (UV)~\cite{peter2014hot,tian2016iris} burst, and campfires~\cite{berghmans2021extreme,alipour2022automatic} is ubiquitous in the lower solar atmosphere, which play important roles in heating the chromosphere and the corona. 
However, the reconnection mechanism in these events is still an open question because of the limited resolution of existing solar telescopes. 
The high plasma density, low temperature and abundant neutral particles indicate that reconnection and the consequent heating in these events might be very different from those in the solar corona.

The role of neutrals in the low atmosphere cannot be ignored in the reconnection process~\cite{ni2021magnetic}. 
Collisions between electrons and neutrals would directly enhance the magnetic diffusion in the reconnection region, which accelerates the reconnection process in the weakly ionized event such as the Ellerman Bomb~\cite{liu2023numerical}. 
The magnetic field in partially ionized plasma could decouple from the center of mass flow when the neutrals become decoupled from the plasma. The decoupling of ions from neutrals begins when the mean free path of the neutral-ion collision exceeds the current sheet thickness, or when the dynamical timescale of the system falls below the collisional timescale. 
Such a decoupling effect between ions and neutrals is evaluated by the ambipolar diffusion terms in the single-fluid MHD approximation.
Due to the decoupling of ions and neutrals caused by ambipolar diffusion, only the ion pressure is available to balance the Lorentz force, which sharpens the magnetic field profile.
Heitsch \& Zweibel~\cite{heitsch2003fast, heitsch2003suppression,brandenburg1994formation} explored the evolution of the current sheet in the presence of ambipolar diffusion, they found that the reconnection rate can be strongly increased by the ambipolar diffusion effect, but even a small guide field could hinder the acceleration of the reconnection rate by the ambipolar diffusion. 
Recent two-fluid MHD simulations confirm that the ion recombination process plays the major role in accelerating the magnetic reconnection process in the low solar atmosphere~\cite{leake2012multi,leake2013magnetic,murtas2021coalescence}.
The effect of radiation on magnetic reconnection in astrophysical and laboratory plasmas were analyzed in many studies~\cite{uzdensky2011magnetic,provornikova2016plasma},
strong radiative losses cool the plasma, which reduces pressure, shrinks the current sheet, and accelerates the reconnection process in the absence of a guide field. 
The high plasma density in the low solar atmosphere indicate that the radiative cooling process should be considered when we study the reconnection mechanism in this region.

We note here as well that neutral particles almost become fully ionized when the temperature in the low $\beta$ reconnection event such as UV bursts reaches above 20, 000 K, then the effects of neutrals vanishes in the process of temperature increase. 
For the first time, Ni et al. 2015~\cite{ni2015fast} showed that the plasmoid instability efficiently accelerates the reconnection process and the ambipolar diffusion is less important in the low $\beta$ magnetic reconnection process in the chromosphere. 
The two-fluids MHD simulation further suggests that the plasmoid instability is the dominant mechanism for fast reconnection in the low $\beta$ environment below the middle chromosphere~\cite{ni2018onset}. 
Recent PIC simulations indicate that fast Hall reconnection becomes efficient in the upper chromosphere~\cite{jara2019kinetic}. 
However, the Hall effect cannot dissipate the magnetic energy and heat the plasmas in the reconnection region.

The highly stratified low solar atmosphere might result in that the reconnection and the heating mechanisms are different in a lower layer from that in a higher one. 
In this work, we numerically study the high $\beta$ magnetic reconnection process in the altitude range from the photosphere to the upper chromosphere. 
The hydrogen-helium plasma with time-dependent ionization degrees are included to generate a more realistic magnetic diffusion as a result of electron-neutral collision, ambipolar diffusion and viscosity, and a more realistic radiative cooling model is also included. 
We focus on looking for the fast reconnection mechanism and the dominant heating mechanisms at different layers of the low solar atmosphere. 
The model is described in Section~\ref{sec_II}. 
In Sections~\ref{sec_III} and~\ref{sec_IV}, the results and discussions about different simulation cases are presented. 
Finally, a summary of the work is given in Section~\ref{sec_V}.

\section{Simulation model}
\label{sec_II}
\subsection{Single-fluid governing equations and important coefficients}
A set of 2.5D simulations are conducted using the single-fluid MHD code, NIRVANA~\cite{ziegler2011semi}. 
In this study, we consider the hydrogen-helium mixture, composed of $H$, $H^{+}$, $H_{e}$, $H_{e}^{+}$ and electrons. 
All the components are strongly coupled and are considered as single fluid. 
A set of single-fluid MHD equations used in our simulations are as follows:
\begin{eqnarray}
\frac{\partial \rho}{\partial t} = - \nabla \cdot (\rho \mathbf{v}),
\label{eq:1}
\end{eqnarray}

\begin{eqnarray}
\begin{split}   
\frac{\partial (\rho \mathbf{v})}{\partial t} & = - \nabla \cdot \left[\rho \mathbf{vv}+\left(p+\frac{1}{2\mu_{0}} |\mathbf{B}|^{2}\right)I-\frac{1}{\mu_{0}}\mathbf{BB}\right] \\
&+\nabla \cdot \tau_{S},
\end{split}
\label{eq:2}
\end{eqnarray}

\begin{eqnarray}
\begin{split}
    \frac{\partial e}{\partial t} &= -\nabla \cdot \left[\left(e+p+\frac{1}{2\mu_{0}} |\mathbf{B}|^{2}\right) \mathbf{v}\right] \\
    & + \nabla \cdot \left[\frac{1}{\mu_{0}} (\mathbf{v} \cdot \mathbf{B})\mathbf{B}\right] \\
    & + \nabla \cdot \left[\mathbf{v} \cdot \tau_{s} + \frac{\eta}{\mu_{0}} \mathbf{B} \times (\nabla \times \mathbf{B})\right] \\
    & - \nabla \cdot \left[\frac{1}{\mu_{0}} \mathbf{B} \times \mathbf{E}_{AD}\right] \\
    & + Q_{rad} +H, 
\end{split}
\label{eq:3}
\end{eqnarray}

\begin{eqnarray}
\frac{\partial \mathbf{B}}{\partial t} = \nabla \times (\mathbf{v} \times \mathbf{B} - \eta \nabla \times \mathbf{B} + \mathbf{E}_{AD}),
\label{eq:4}
\end{eqnarray}
with
\begin{eqnarray}
e = \frac{p}{\gamma -1} + \frac{1}{2} \rho |\mathbf{v}|^{2} + \frac{1}{2 \mu_{0}} |\mathbf{B}|^{2}
\label{eq:5}
\end{eqnarray}
and
\begin{eqnarray}
p = \frac{(1.1+Y_{iH}+0.1Y_{iHe})\rho}{1.4 m_{i}} k_{B} T,
\label{eq:6}
\end{eqnarray}
where $\rho$ is the plasma mass density, $\mathbf{v}$ is the fluid velocity, $p$ is the plasma thermal pressure, $\mathbf{B}$ is the magnetic field, $e$ is the total energy density, $T$ is the temperature; and $Y_{iH}$ and $Y_{iHe}$ are the hydrogen and helium ionization fractions, respectively, while m$_{i}$ is the mass of proton, and $k_{B}$ is the Boltzmann constant. 
The total helium number density with respect to that of the hydrogen is set to 10\%, only the primary ionization of helium is considered. 
 The adiabatic constant $\gamma$ is set to 5/3. The stress tensor is $\tau_{S} = \xi [\nabla \mathbf{v} + (\nabla \mathbf{v})^{T} - \frac{2}{3} (\nabla \cdot \mathbf{v})I]$ , where $\xi$ represents the dynamic viscosity coefficient in the units of kg m$^{-1}$ s$^{-1}$. As the current sheet studied in this work is aligned parallel to the solar surface, we ignore the gravity, and the initial plasma density is assumed to be constant in the whole domain.

Interactions of various species in the partially ionized plasma in the low solar atmosphere are of great interest. 
The magnetic diffusion ($\eta$) caused by electron collisions with ions and neutrals is given as~\cite{khomenko2012heating,ni2022plausibility}    
\begin{eqnarray}
\eta = \eta_{ei} + \eta_{en} = \frac{m_{e} \nu_{ei}}{e^{2}_{c} n_{e} \mu_{0}} + \frac{m_{e} \nu_{en}}{e^{2}_{c} n_{e} \mu_{0}}
\label{eq:7}
\end{eqnarray}
where $m_{e}$, $e_{c}$, $\mu_{0}$, $n_{e}$, $\nu_{en}$, $\nu_{ei}$ are the electron mass, electron charge, permeability of vacuum, electron density, collision frequency of electron-neutral and electron-ion, respectively. The electron number density and the collision frequencies read~\cite{ni2022plausibility}  

\begin{eqnarray}
n_{e} = \frac{\rho(Y_{iH}+0.1Y_{iHe})}{1.4m_{i}} \text{ and } \nu_{ei} = \frac{n_{e} e^{4}_{c} \Lambda}{3 m^{2}_{e} \epsilon^{2}_{0}} \left( \frac{m_{e}}{2 \pi k_{B} T} \right)^{3/2}, 
\label{eq:8}
\end{eqnarray}
respectively. We then further have~\cite{ni2022plausibility}
\begin{eqnarray}
\nu_{en} = n_{n} \sqrt{\frac{8 k_{B} T}{\pi m_{en}}} \sigma_{en}.
\label{eq:10}
\end{eqnarray}
Here $\Lambda$ is the Coulomb logarithm, $\epsilon_{0}$ is the permittivity of free space, $n_{n}$ is the number density of neutrals, and $\sigma_{en}$ is the collisional cross-section. The value of $\Lambda$ is determined by~\cite{ni2022plausibility}
\begin{eqnarray}
\Lambda = 23.4-1.15 \log_{10} n_{e} + 3.45 \log_{10} T.
\label{eq:11}
\end{eqnarray}
For the hydrogen-helium plasma, the electron-neutral collision frequency ($\nu_{en}$) are due to collisions of electrons with both neutral helium and neutral hydrogen. The electron-neutral collision frequency is

\begin{eqnarray}
\nu_{en} = n_{n H_{e}} \sqrt{\frac{8 k_{B} T}{\pi m_{e}}} \sigma_{e-n H_{e}} + n_{n H} \sqrt{\frac{8 k_{B} T}{\pi m_{e}}} \sigma_{e-n H},
\label{eq:12}
\end{eqnarray}
where $n_{nHe} = 0.1\rho(1-Y_{iHe})/(1.4mi)$ is the number density of the neutral helium and $n_{nH} = \rho(1-Y_{iH})/(1.4mi)$ is the number density of the neutral hydrogen. The electron-neutral hydrogen collision cross-section ($\sigma_{e-n H}$) and electron-neutral helium collision cross-sections ($\sigma_{e-n H_{e}}$) values are $2 \times 10^{-19}$ m$^{2}$ and $\sigma_{e-n H}/3$, respectively~\cite{vranjes2013collisions}. 
Substituting Eqs.~(\ref{eq:8})-(\ref{eq:12}) into Eq.~(\ref{eq:7}) gives the magnetic diffusivities
\begin{eqnarray}
\eta_{ei} \simeq 1.0246 \times 10^{8} \Lambda T^{-1.5}
\label{eq:13}
\end{eqnarray}
for the electron-ion collisions, and
\begin{eqnarray}
\eta_{en} \simeq 0.0351 \sqrt{T} \frac{\left[\frac{0.1}{3} (1-Y_{iH_{e}})+(1-Y_{iH})\right]}{Y_{iH} + 0.1 Y_{iHe}}
\label{eq:14}
\end{eqnarray}
for the electron-neutral collisions, where $\eta_{ei}$ and $\eta_{en}$ are in m$^{2}$ s$^{-1}$.

The ambipolar electric field (E$_{AD}$) included in the energy and the induction equations~(\ref{eq:3}) and (\ref{eq:4}) is~\cite{ni2021magnetic,ni2022plausibility}
\begin{eqnarray}
\mathbf{E}_{AD} = \frac{1}{\mu_{0}} \eta_{AD} [(\nabla \times B) \times B ] \times B,
\label{eq:15}
\end{eqnarray}
where $\eta_{AD}$ is the coefficient of the ambipolar diffusion such as~\cite{khomenko2012heating,ni2020magnetic,ni2022plausibility} 
\begin{eqnarray}
\eta_{AD} = \frac{(\rho_{n}/\rho)^2}{\rho_{i} \nu_{in} + \rho_{e} \nu_{en}}
\label{eq:16}
\end{eqnarray}
in units of m$^{3}$ s kg$^{-1}$. Since we are dealing with plasma containing both helium and hydrogen, the $\rho_{n}/\rho$ is given by~\cite{ni2022plausibility}
\begin{eqnarray}
\rho_{n}/\rho  = \frac{0.4 (1- Y_{iHe}) + (1- Y_{iH})}{1.4}.
\label{eq:17}
\end{eqnarray}
The ion collision part reads~\cite{ni2022plausibility}
\begin{eqnarray}
\begin{split}
    \rho_{i} \nu_{in} & = \rho_{iH} n_{nH} \sqrt{\frac{8 k_{B} T}{\pi m_{i}/2}} \sigma_{iH-nH} \\
    & + \rho_{iH} n_{nHe} \sqrt{\frac{8 k_{B} T}{4 \pi m_{i}/5}} \sigma_{iH-nHe} \\
    & + \rho_{iHe} n_{nH} \sqrt{\frac{8 k_{B} T}{4 \pi m_{i}/5}} \sigma_{iHe-nH} \\
    & + \rho_{iHe} n_{nHe} \sqrt{\frac{8 k_{B} T}{2 \pi m_{i}}} \sigma_{iHe-nHe}, 
\end{split}
\label{eq:18}
\end{eqnarray}
where $\rho_{iH} = \rho Y_{iH}/1.4$ and $\rho_{iHe} = 0.4\rho Y_{iHe}/1.4$ are the ionized hydrogen and the helium mass densities, respectively; $\sigma_{iH-nH}$ is the ionized hydrogen-neutral hydrogen collisional cross-section, $\sigma_{iH-nHe}$ is the collisional cross-section of ionized hydrogen-neutral helium, $\sigma_{iHe-nH}$ is that of the ionized helium-neutral hydrogen collision, and  $\sigma_{iHe-nHe}$ is that of the ionized helium-neutral helium collision. We choose $\sigma_{iH-nH} = 1.5 \times 10^{-18}$ m$^{2}$, $\sigma_{iH-nHe}$ = $\sigma_{iHe-nH}$ = $\sigma_{iHe-nHe}$ = $\sigma_{iH-nH}/\sqrt{3}$ given by Varnjes \& Krstic~\cite{vranjes2013collisions} and Barata \& Conde~\cite{barata2010elastic}. 
The electron collision contribution part is written as~\cite{ni2022plausibility} 
\begin{eqnarray}
\begin{split}
    \rho_{e} \nu_{en} & = \rho_{e} n_{nH} \sqrt{\frac{8 k_{B} T}{\pi m_{e}}} \sigma_{e-nH} \\
    & + \rho_{e} n_{nHe} \sqrt{\frac{8 k_{B} T}{\pi m_{e}}} \sigma_{e-nHe}, \\
\end{split}
\label{eq:19}
\end{eqnarray}
where $\sigma_{e-nH, nHe}$ refers to the collisional cross-sections of electron-neutral hydrogen and electron-neutral helium, respectively. Both these collisional cross-sections are smaller than the collisional cross-section contributed by ions and neutrals ($\sigma_{e-nHe} < \sigma_{e-nH} < \sigma_{iH-nH}$), therefore electron collisions are ignored, and Eq.~(\ref{eq:16}) is simplified into $\eta_{AD} = \frac{(\rho_{n}/\rho)^2}{\rho_{i} \nu_{in}}$.

The dynamic viscosity due to the contribution of ions and neutrals in partially ionized plasma is given by~\cite{ni2022plausibility}   
\begin{eqnarray}
\xi = \xi_{i} + \xi_{n} = \frac{n_{i} k_{B} T}{\nu_{ii}} + \frac{n_{n} k_{B} T}{\nu_{nn}},
\label{eq:20}
\end{eqnarray}
where $\xi_{i,n}$ represents the ion and the neutral viscosity coefficient, and $\nu_{nn,ii}$ is the collision frequencies between neutral-neutral and ion-ion, the collision frequencies such as~\cite{leake2013magnetic}: 
\begin{eqnarray}
\nu_{nn} = n_{n} \sigma_{nn} \sqrt{\frac{16 k_{B} T}{\pi m_{n}}}
\label{eq:21}
\end{eqnarray}
and
\begin{eqnarray}
\nu_{ii} = \frac{n_{i} e^{4}_{c} \Lambda}{3 m^{2}_{i} \epsilon^{2}_{0}} \left(\frac{m_{i}}{2 \pi k_{B} T} \right)^{3/2}.
\label{eq:22}
\end{eqnarray}
In this work, we considered the contributions of both hydrogen and helium components to get a more realistic viscosity than those in previous work~\cite{ni2022plausibility}, which is
\begin{eqnarray}
\xi = \xi_{i} + \xi_{n} \simeq  \frac{4.8692 \times 10^{-16}}{\Lambda} T^{2} \sqrt{T}+ 2.0127 \times 10^{-7} \sqrt{T}. 
\label{eq:24}
\end{eqnarray}

As both the coronal equilibrium and the Saha ionization equilibrium are not applicable in the solar chromosphere, the temperature dependent ionization degree of hydrogen (Y$_{iH}$) and helium (Y$_{iHe}$) based on the RADYN test atmosphere results by solving the radiative transfer equations~\cite{carlsson2012approximations} are used in this work. 
We refer readers to Figure 1 of Ni et al. 2022~\cite{ni2022plausibility} and Liu et al. 2023~\cite{liu2023numerical}, where one can see the variation of hydrogen and helium ionization degrees with plasma temperature.

\subsection{Radiation models}
Radiation in the low solar atmosphere is strong. Therefore, it is important to analyze the effect of radiative cooling on the magnetic reconnection process.
Three different radiative cooling models are considered in this study.
In order to include effective radiative cooling process for photospheric reconnection, the models of Gan \& Fang~\cite{gan1990hydrodynamic} and Abbett \& Fisher~\cite{abbett2012radiative} are applied in the magnetic reconnection process inside the photosphere. Whereas, the Carlsson \& Leenaarts~\cite{carlsson2012approximations} model is used for the reconnection process in the chromosphere.
The Gan \& Fang radiative cooling model, which is applicable up to T $\sim$ 10$^{5}$ K, reads as 
\begin{eqnarray}
Q_{rad1} = - 1.547 \times 10^{-42} n_{e} n_{H} \alpha T^{1.5},
\label{eq:25}
\end{eqnarray}
where n$_{e}$, n$_{H}$ are number densities of electrons and hydrogen, respectively. The electron density is deduced by using the modified Saha and Boltzamann expression~\cite{gan1990hydrodynamic,fang2002magnetic}, and  $\alpha$ is a function of the height and is approximately equal to 6.558 $\times$ 10$^{-5}$ at 400 km above the solar surface.
The $\alpha$ is~\cite{gan1990hydrodynamic,cheng2021ellerman}
\begin{eqnarray}
\begin{split}    
\alpha &= 10^{a1} + 2.3738\times 10^{-4}e^{a2},\\
a1 &= 2.75 \times 10^{-3}Z-5.445, \\
a2 &= \frac{-Z}{163}
\label{eq:25a}
\end{split}
\end{eqnarray}
where $Z$ is height in km.

The Abbett \& Fisher model used in the photospheric magnetic reconnection is
\begin{eqnarray}
Q_{rad2} = - 2 \kappa^{B} \rho \sigma T^{4} E(\tau^{B}).
\label{eq:26}
\end{eqnarray}
where $\kappa^{B}$ is the mean opacity which depends on plasma temperature and density, $\tau^{B}$ is the optical depth computed from the mean opacity, and $\sigma$ is the Stefan-Boltzmann constant. The table about the relationship between $\tau^{B}$, temperature and density is used from Iglesias \& Rogers~\cite{iglesias1996updated}.  
We are using Rosseland mean opacity instead of Planck-weighted mean opacity because both values are similar at relevant photospheric altitudes.
With this scheme, computation of the radiative heating only requires a simple 2D table lookup to get $\kappa^{B}$ and a column-by-column integration over depth to compute $\tau^{B}$. The values of $\kappa^{B}$ and $\tau^{B}$ at 400 km are $3.197 \times 10^{-10}$ and $1.499 \times 10^{-3}$, respectively.

The Carlsson \& Leenaarts model is based on a small number of strong lines from neutral hydrogen, singly ionized calcium and singly ionized magnesium. The model reads

\begin{eqnarray}
Q_{rad3} = - \sum_{X = H, Mg, Ca} L_{Xm} (T) E_{Xm} (\tau) \frac{N_{Xm}}{N_{X}} (T) A_{X} \frac{N_{H}}{\rho} n_{e} \rho,
\label{eq:27}
\end{eqnarray}
where $L_{Xm} (T)$ is the optically thin radiative loss function depends on temperature $T$ , per electron and per particle of element $X$ in ionization stage $m$, 
$E_{Xm} (\tau)$ is the escape probability varying with the optical depth $\tau$, 
$\frac{N_{Xm}}{N_{X}}(T)$ is the fraction of element $X$ which is in ionization stage $m$, 
$A_{X}$ is the abundance of element $X$, 
and $\frac{N_{H}}{\rho} = 4.407 \times 10^{23} g^{-1}$ is the number of hydrogen
per unit mass of the chromospheric material. 
The optical depth ($\tau$) used in eq.~\ref{eq:27} is computed by multiplying the total column density of neutral hydrogen with the constant $4.0 \times 10^{-14}$ cm$^{2}$.
For hydrogen, $L_{Xm}$, $E_{Xm}$, and $N_{Xm}/N_{X}$ are obtained from a 1D radiation hydrodynamic simulation including non-equilibrium ionization computed using the RADYN code.
Whereas, for Mg and Ca such quantities were computed from a 2D radiation-MHD simulation with BIFROST, which provided the atmospheric structure and radiative transfer calculations using MULTI3D~\cite{leenaarts2009multi3d}.

\subsection{Simulation setup}
In this study, MHD simulations are performed to investigate the evolution of the current sheet using the photospheric and the chromoshperic plasma conditions. 
The same setup in all the cases are described as follows. 
The size of simulation domain extends from 0 to $L_{0}$ and $-0.5L_{0}$ to $0.5L_{0}$ in the $x$ and $y$ directions, respectively. Where the length scale $L_{0}$ = 2 $\times$ 10$^{5}$ m, which is comparable with the length scale of small scale reconnection events observed in the photosphere and the chromosphere. Open boundary conditions are used in both directions.
The adaptive mesh refinement (AMR) technique is applied in the present work, which starts with a base-level grid of 192 $\times$ 192 and the highest AMR level is 9. The initial plasma temperature (T$_{0}$) and mass density ($\rho_{0}$) as a function of height (Z) are obtained from C7 atmosphere model. There are no exact explicit functions of how T$_{0}$ and $\rho_{0}$ depends on Z, we get the values of T$_{0}$ and $\rho_{0}$ at different heights from table 26 of Avrett \& Loeser 2008~\cite{avrett2008models}. We have investigated 13 cases, the key initial parameters for each case are listed in Table~\ref{tab:table1}.
The simulations are initialized with the horizontal force-free Harris sheet in equilibrium, which is given by:

\begin{eqnarray}
B_{x0} = - b_{0} \tanh [y/(0.05L_{0})],
\label{eq:28}
\end{eqnarray}
\begin{eqnarray}
B_{y0} = 0,
\label{eq:29}
\end{eqnarray}
\begin{eqnarray}
B_{z0} = b_{0}/ \cosh [y/(0.05L_{0})],
\label{eq:30}
\end{eqnarray}
where $b_{0}$ is the initial magnetic field. The small magnetic perturbation applied at the beginning of simulation to trigger magnetic reconnection is given below:
\begin{eqnarray}
b_{x1} = - b_{pert} \sin \left[\frac{2 \pi (y+0.5L_{0})}{L_{0}}\right] \cos \left[\frac{2 \pi (x+0.5L_{0})}{L_{0}}\right],
\label{eq:31}
\end{eqnarray}

\begin{eqnarray}
b_{y1} = b_{pert} \cos \left[\frac{2 \pi (y+0.5L_{0})}{L_{0}}\right] \sin \left[\frac{2 \pi (x+0.5L_{0})}{L_{0}}\right]
\label{eq:32}
\end{eqnarray}
with b$_{pert}$  = 0.005 b$_{0}$.
\begin{figure}[htbp] 
\centering
\begin{minipage}{0.495\textwidth}
\includegraphics[width=0.99\textwidth]{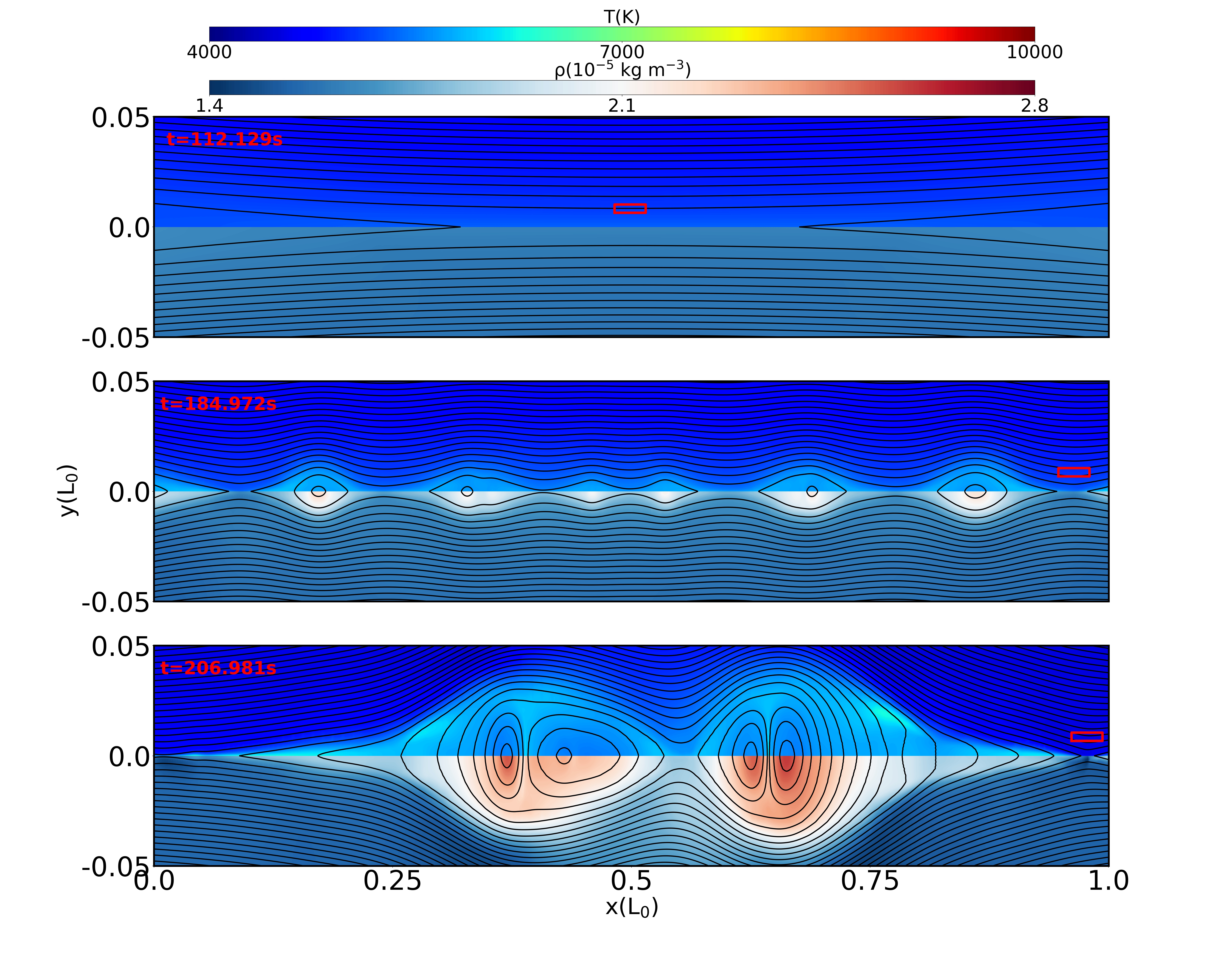}
\put(-160,158){\textbf{(a) Case-I (Z = 400 km, Q$_{rad1}$)}}
\end{minipage}
\begin{minipage}{0.495\textwidth}
\includegraphics[width=0.99\textwidth]{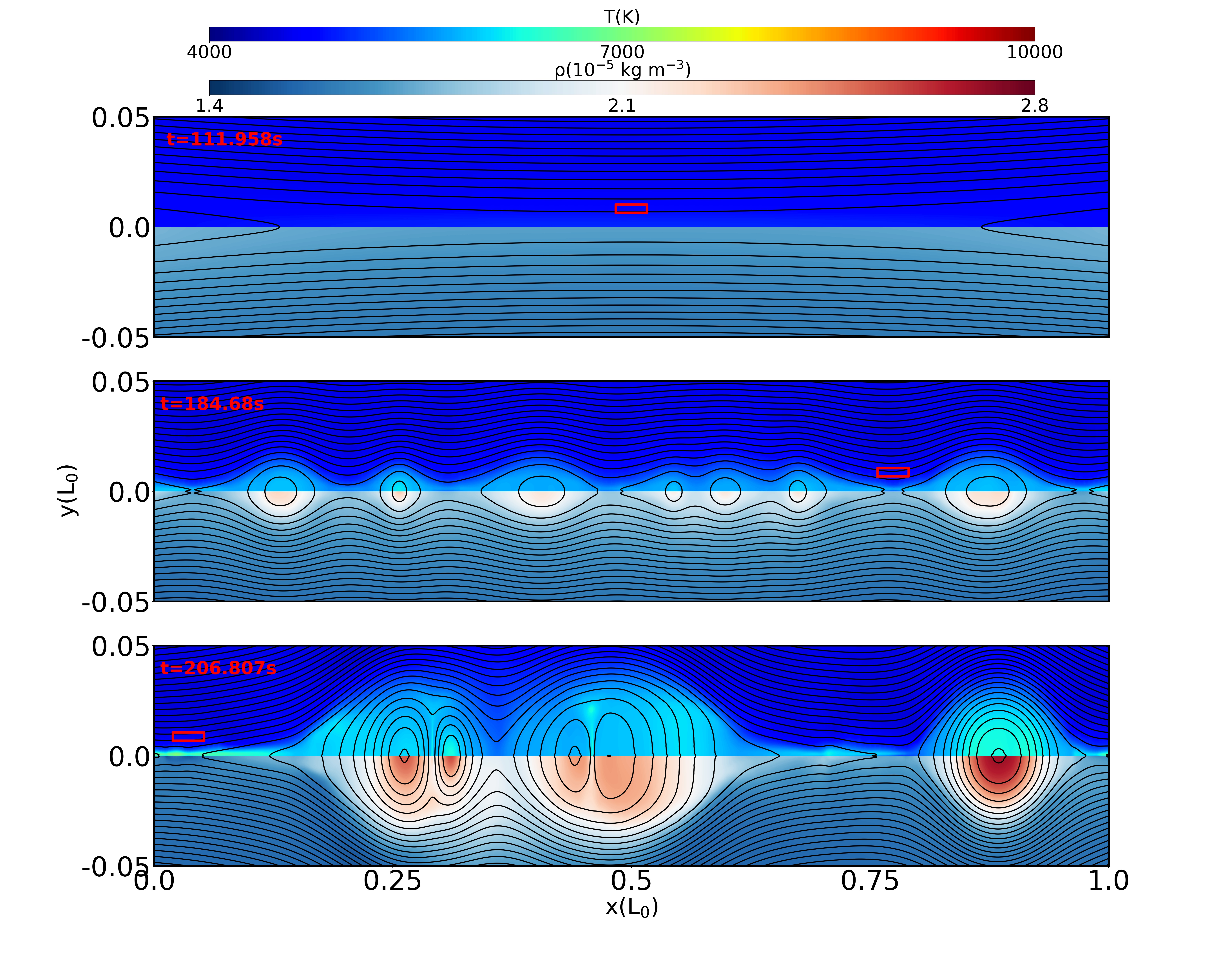}
\put(-160,158){\textbf{(b) Case-II (Z = 400 km, Q$_{rad2}$)}}
\end{minipage}

\vspace{1.0mm}
\begin{minipage}{0.495\textwidth}
\includegraphics[width=0.99\textwidth]{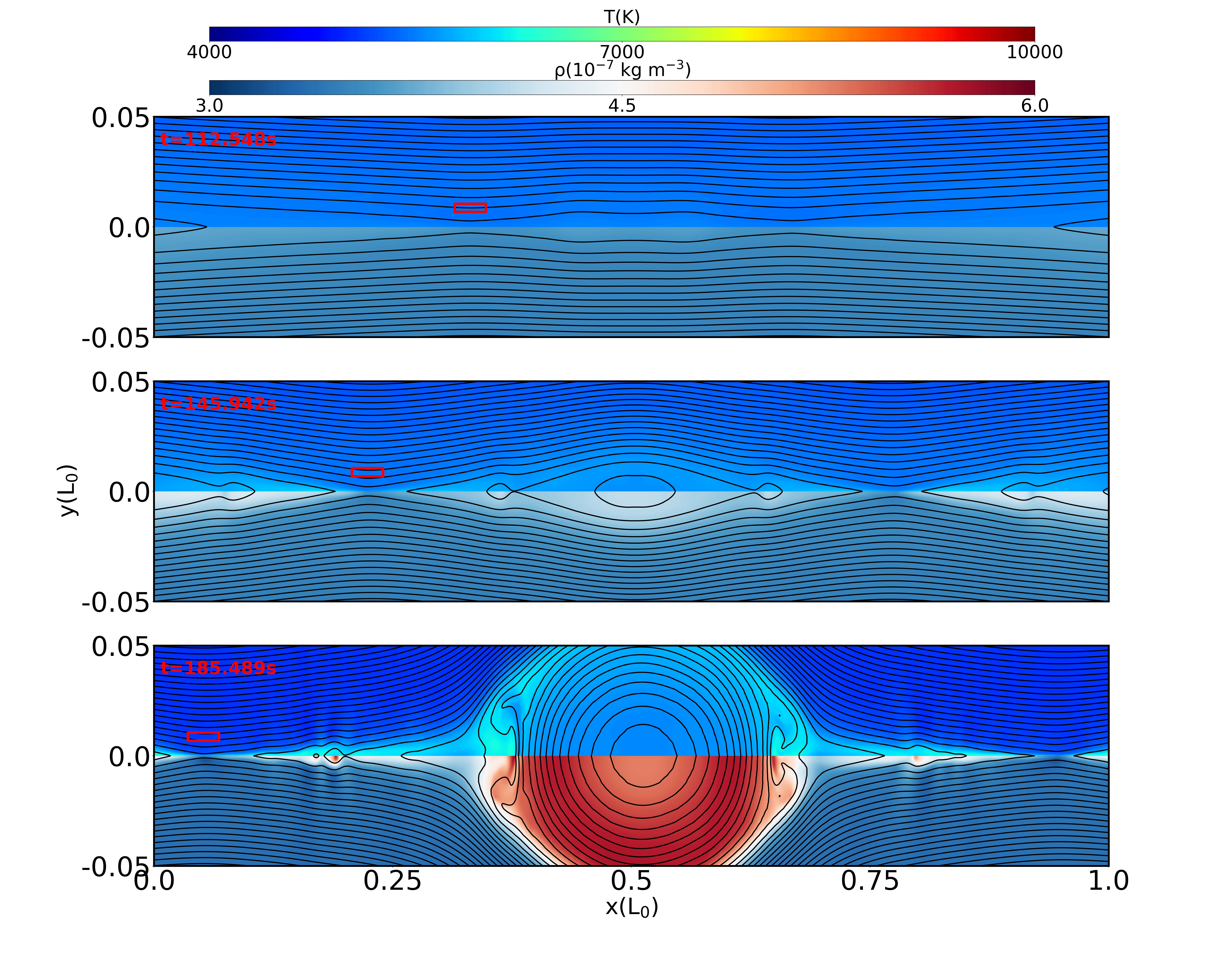}
\put(-160,158){\textbf{(c) Case-III (Z = 800 km, Q$_{rad3}$)}}
\end{minipage}
\begin{minipage}{0.495\textwidth}
\includegraphics[width=0.99\textwidth]{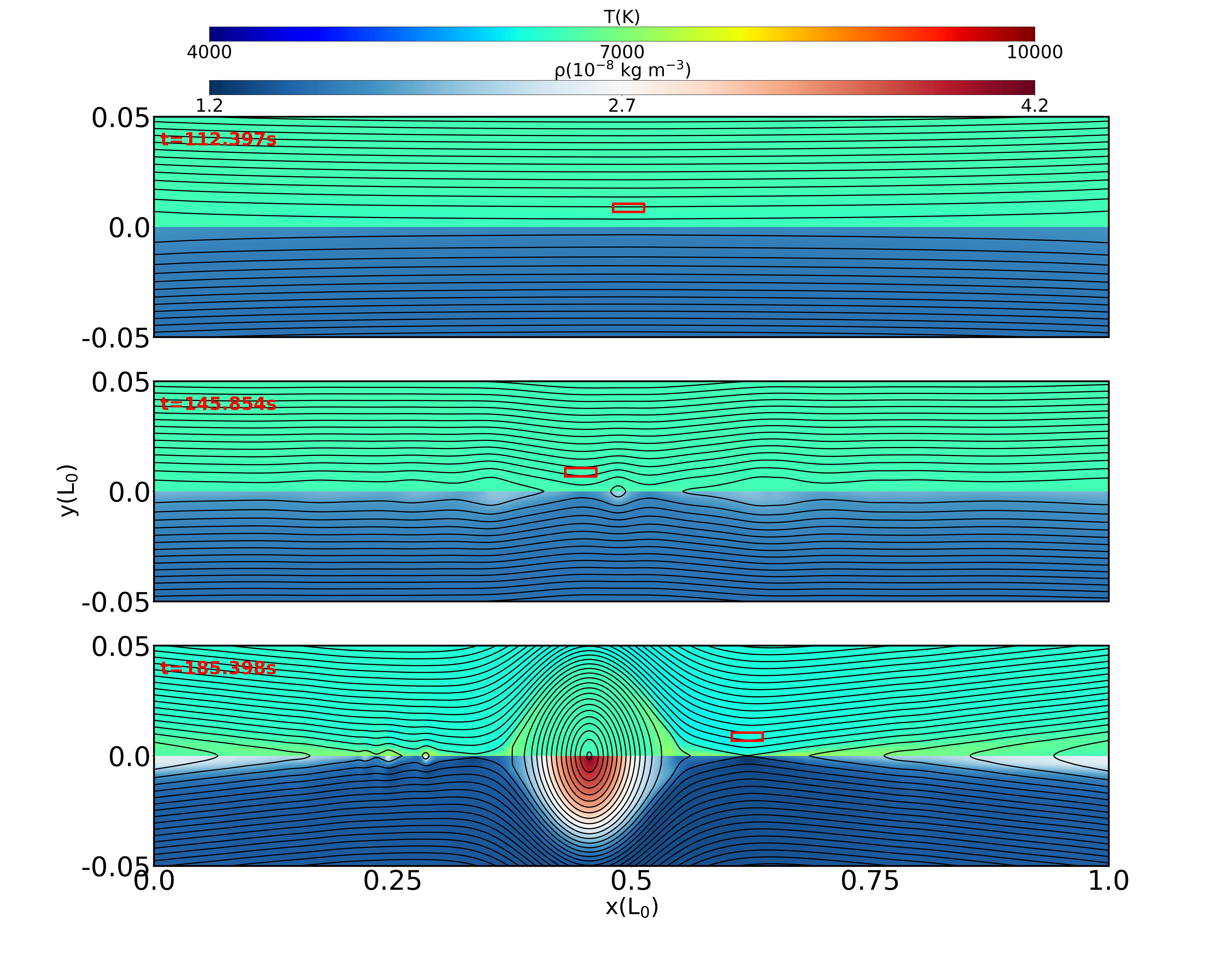}
\put(-160,158){\textbf{(d) Case-IV (Z = 1200 km, Q$_{rad3}$)}}
\end{minipage}

\vspace{1.0mm}
\begin{minipage}{0.495\textwidth}
\includegraphics[width=0.99\textwidth]{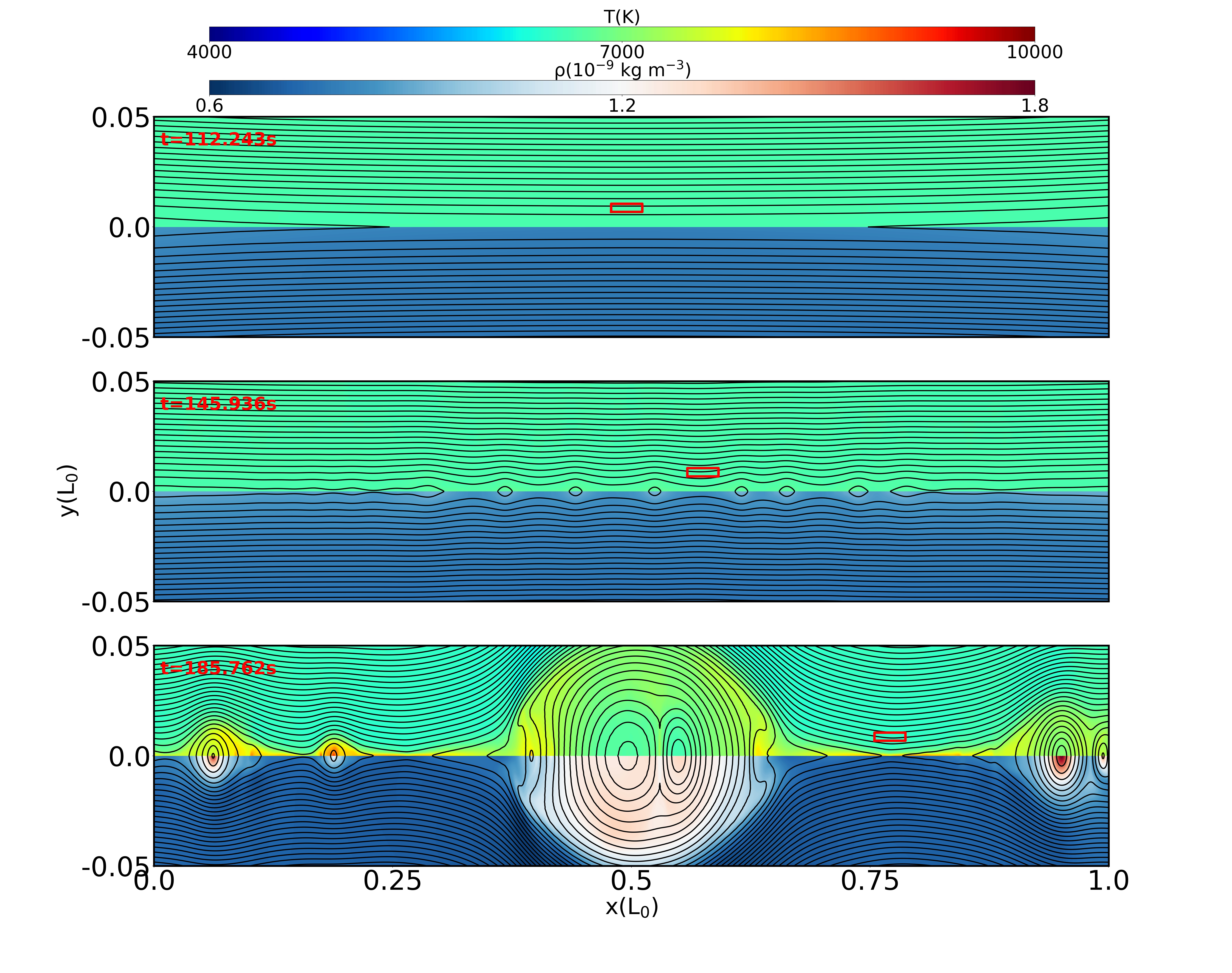}
\put(-160,158){\textbf{(e) Case-V (Z = 1700 km, Q$_{rad3}$)}}
\end{minipage}
\begin{minipage}{0.495\textwidth}
\includegraphics[width=0.99\textwidth]{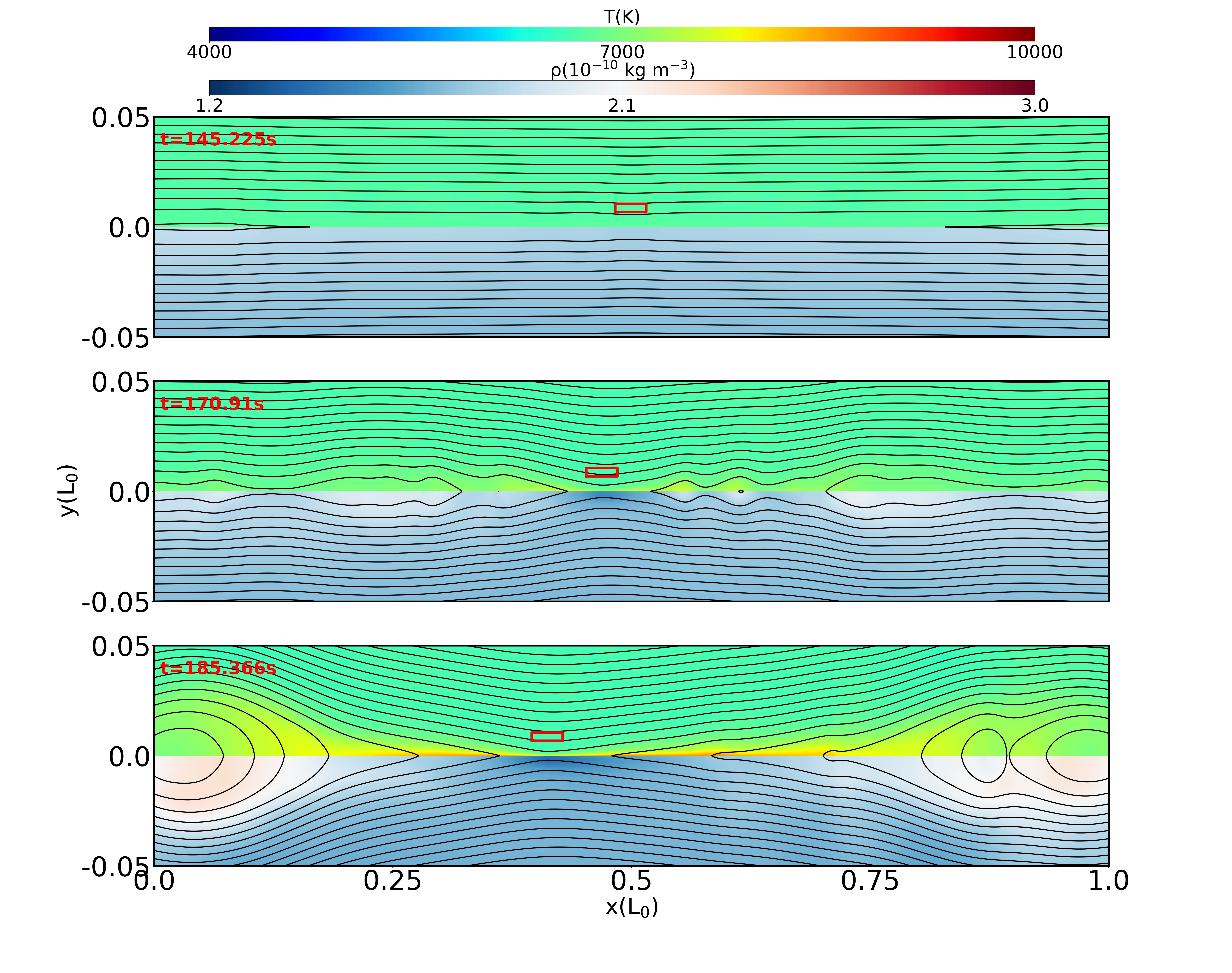}
\put(-160,158){\textbf{(f) Case-V (Z = 2000 km, Q$_{rad3}$)}}
\end{minipage}
\caption{The 2D distributions of the temperature and plasma density at three typical times in six different cases. The two different background colors in half of the upper and the lower panels represent the temperature and the density, respectively. The black solid lines represent the magnetic field lines. The red boxes are the regions for calculating the reconnection rates.}
\label{fig_1}
\end{figure}



\begin{table}
\caption{\label{tab:table1} Input parameters used in the simulations.}
  \centerline{%
  \resizebox{0.99\textwidth}{!}{
    \begin{tabular}{|l|*9{c|} l|}
      \hline
    \cline{2 - 11}
    & Z (km) & Radiative Cooling & Ambipolar Diffusion & Magnetic Diffusion & Viscosity & $\rho_{0}$ (kg m$^{-3}$) & T$_0$ (K)& b$_0$ (T) & Lundquist number (S$_{0}$) & location \\
    \hline
    Case-I & 400 & Q$_{rad1}$ & ON & $\eta_{ei}$ + $\eta_{en}$ & ON & 1.56 $\times$ 10$^{-5}$ & 4590 & 2.98 $\times$ 10$^{-2}$ & 1.840 $\times$ 10$^{4}$ & photosphere\\
    \hline
    Case-II & 400 & Q$_{rad2}$ & ON & $\eta_{ei}$ + $\eta_{en}$ & ON & 1.56 $\times$ 10$^{-5}$ & 4590 & 2.98 $\times$ 10$^{-2}$ & 1.840 $\times$ 10$^{4}$ & photosphere\\
    \hline
    Case-III & 800 & Q$_{rad3}$ & ON & $\eta_{ei}$ + $\eta_{en}$ & ON & 3.44 $\times$ 10$^{-7}$ & 5100 & 4.66 $\times$ 10$^{-3}$ & 1.841 $\times$ 10$^{4}$ & lower chromosphere\\
    \hline
    Case-IV & 1200 & Q$_{rad3}$ & ON & $\eta_{ei}$ + $\eta_{en}$ & ON & 1.56 $\times$ 10$^{-8}$ & 6574 & 1.16 $\times$ 10$^{-3}$ & 9.253 $\times$ 10$^{5}$ & middle chromosphere\\
    \hline
    Case-V & 1700 & Q$_{rad3}$ & ON & $\eta_{ei}$ + $\eta_{en}$ & ON & 7.44 $\times$ 10$^{-10}$ & 6641 & 2.50 $\times$ 10$^{-4}$ & 8.234 $\times$ 10$^{5}$ & upper chromosphere\\
    \hline
    Case-VI & 2000 & Q$_{rad3}$ & ON & $\eta_{ei}$ + $\eta_{en}$ & ON & 1.68 $\times$ 10$^{-10}$ & 6678 & 1.22 $\times$ 10$^{-4}$ & 7.842 $\times$ 10$^{5}$ & upper chromosphere\\
    \hline
    Case-VII & 400 & OFF & ON & $\eta_{ei}$ + $\eta_{en}$ & ON & 1.56 $\times$ 10$^{-5}$ & 4590 & 2.98 $\times$ 10$^{-2}$ & 1.840 $\times$ 10$^{4}$ & photosphere\\
    \hline
    Case-VIII & 800 & OFF & ON & $\eta_{ei}$ + $\eta_{en}$ & ON & 3.44 $\times$ 10$^{-7}$ & 5100 & 4.66 $\times$ 10$^{-3}$ & 1.841 $\times$ 10$^{4}$ & lower chromosphere\\
    \hline
    Case-IX & 2000 & Q$_{rad3}$ & OFF & $\eta_{ei}$ + $\eta_{en}$ & ON & 1.68 $\times$ 10$^{-10}$ & 6678 & 1.22 $\times$ 10$^{-4}$ & 7.842 $\times$ 10$^{5}$ & upper chromosphere\\
    \hline
    Case-X & 2000 & Q$_{rad3}$ & ON & $\eta_{ei}$ + $\eta_{en}$ & OFF & 1.68 $\times$ 10$^{-10}$ & 6678 & 1.22 $\times$ 10$^{-4}$ & 7.842 $\times$ 10$^{5}$ & upper chromosphere\\
    \hline
    Case-XI & 2000 & OFF & ON & $\eta_{ei}$ + $\eta_{en}$ & ON & 1.68 $\times$ 10$^{-10}$ & 6678 & 1.22 $\times$ 10$^{-4}$ & 7.842 $\times$ 10$^{5}$ & upper chromosphere\\
    \hline
    Case-XII & 400 & Q$_{rad1}$ & ON & $\eta_{ei}$ & ON & 1.56 $\times$ 10$^{-5}$ & 4590 & 2.98 $\times$ 10$^{-2}$ & 1.840 $\times$ 10$^{4}$ & photosphere\\
    \hline
    Case-XIII & 800 & Q$_{rad3}$ & ON & $\eta_{ei}$ & ON & 3.44 $\times$ 10$^{-7}$ & 5100 & 4.66 $\times$ 10$^{-3}$ & 1.841 $\times$ 10$^{4}$ & lower chromosphere\\
    \hline
    \end{tabular}%
    }
   }
\end{table}


\section{Numerical Results} 
\label{sec_III}

The magnetic reconnection processes at five different heights in the low solar atmosphere are described below.
First, We performed magnetic reconnection experiments in the photosphere at Z = 400 km above the solar surface using two different radiative cooling models ($Q_{rad1}$ and $Q_{rad2}$). 
Second, magnetic reconnection is investigated at four different chromospheric altitudes, from the bottom to the top, using the same radiative cooling mode (Q$_{rad3}$).  
The same initial plasma-$\beta$ ($\beta_{0}$ = 1.33) is used in all simulations, and the other important initial parameters are listed in Table~\ref{tab:table1}.
The initial Lundquist number is defined by $S_{0} = L_{0} V_{A0}/\eta$, with $V_{A0}$ the Alfven velocity, $V_{A0} = b_{0}/\sqrt{\mu_{0}\rho_{0}}$ and $\eta$ the initial magnetic diffusivity, which is the sum of diffusion caused by electron-ion collisions ($\eta_{ei}$) and electron-neutral collisions ($\eta_{en}$). 
The Lundquist number increases during the reconnection process due to the decrease of magnetic diffusivity at the main X-point.

\begin{figure}[htbp] 
\centering
\begin{minipage}{0.495\textwidth}
\includegraphics[width=1.0\textwidth]{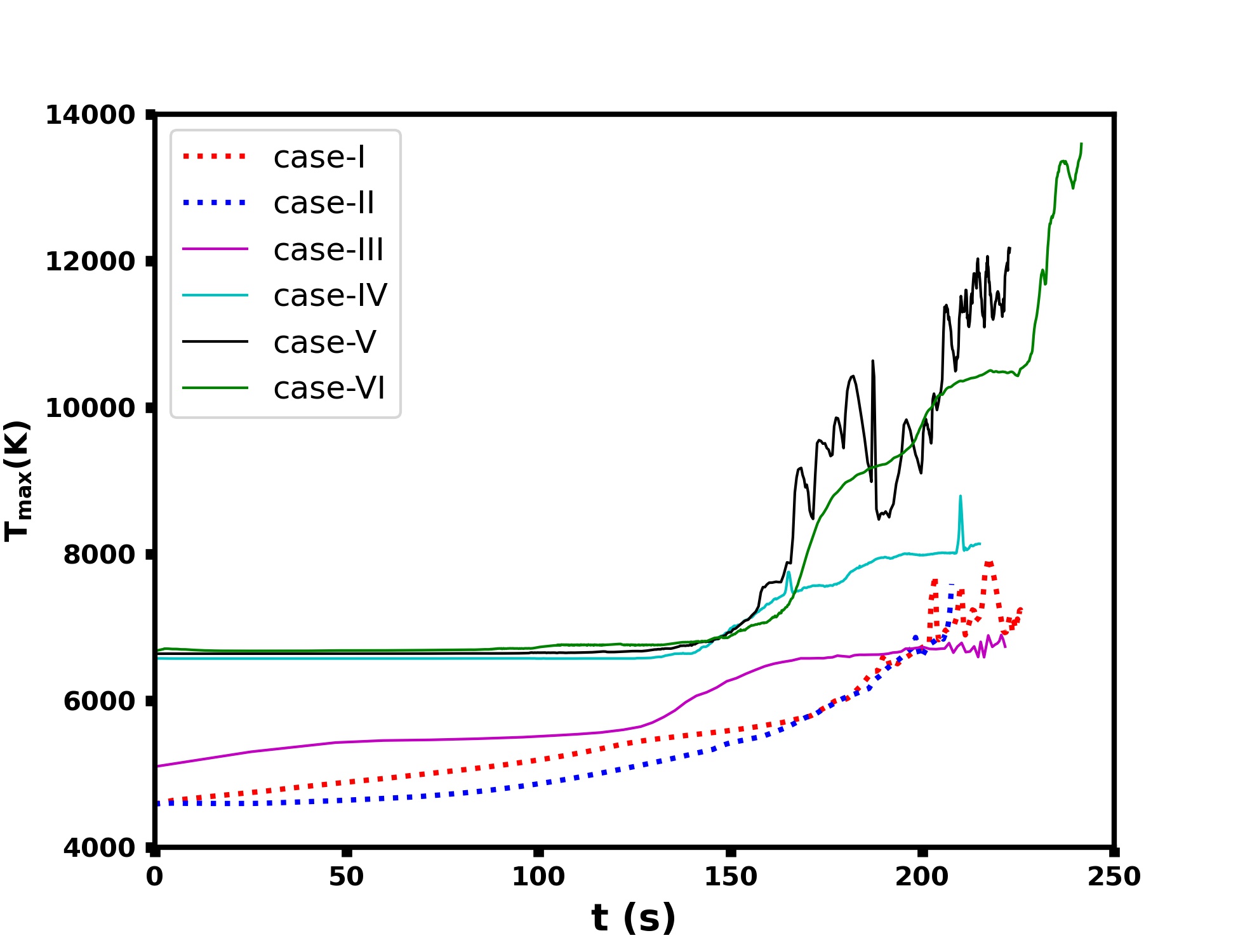}
\put(-110,140){\textbf{(a)}}
\end{minipage}
\begin{minipage}{0.495\textwidth}
\includegraphics[width=1.0\textwidth]{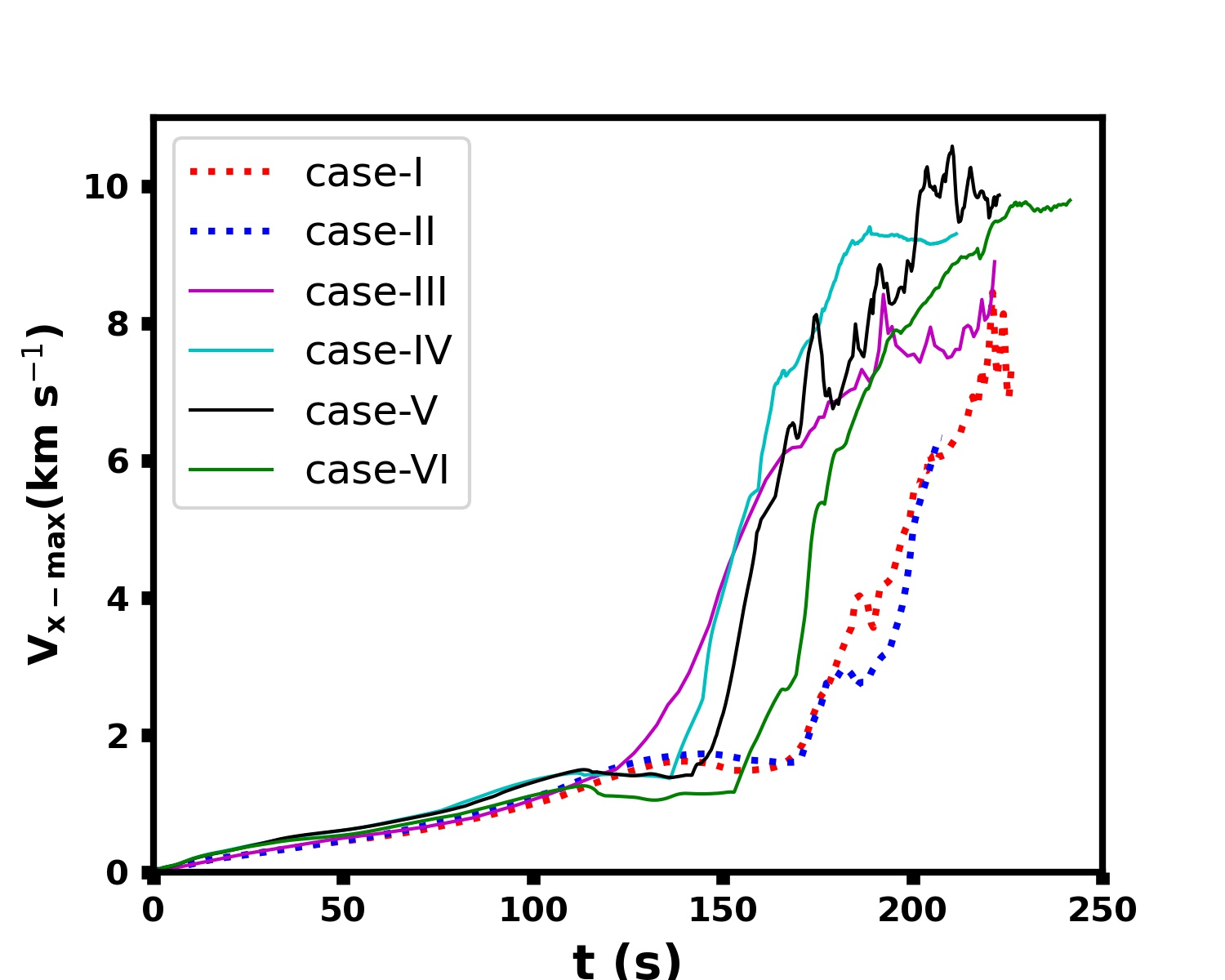}
\put(-110,145){\textbf{(b)}}
\end{minipage}
\begin{minipage}{0.495\textwidth}
\includegraphics[width=1.0\textwidth]{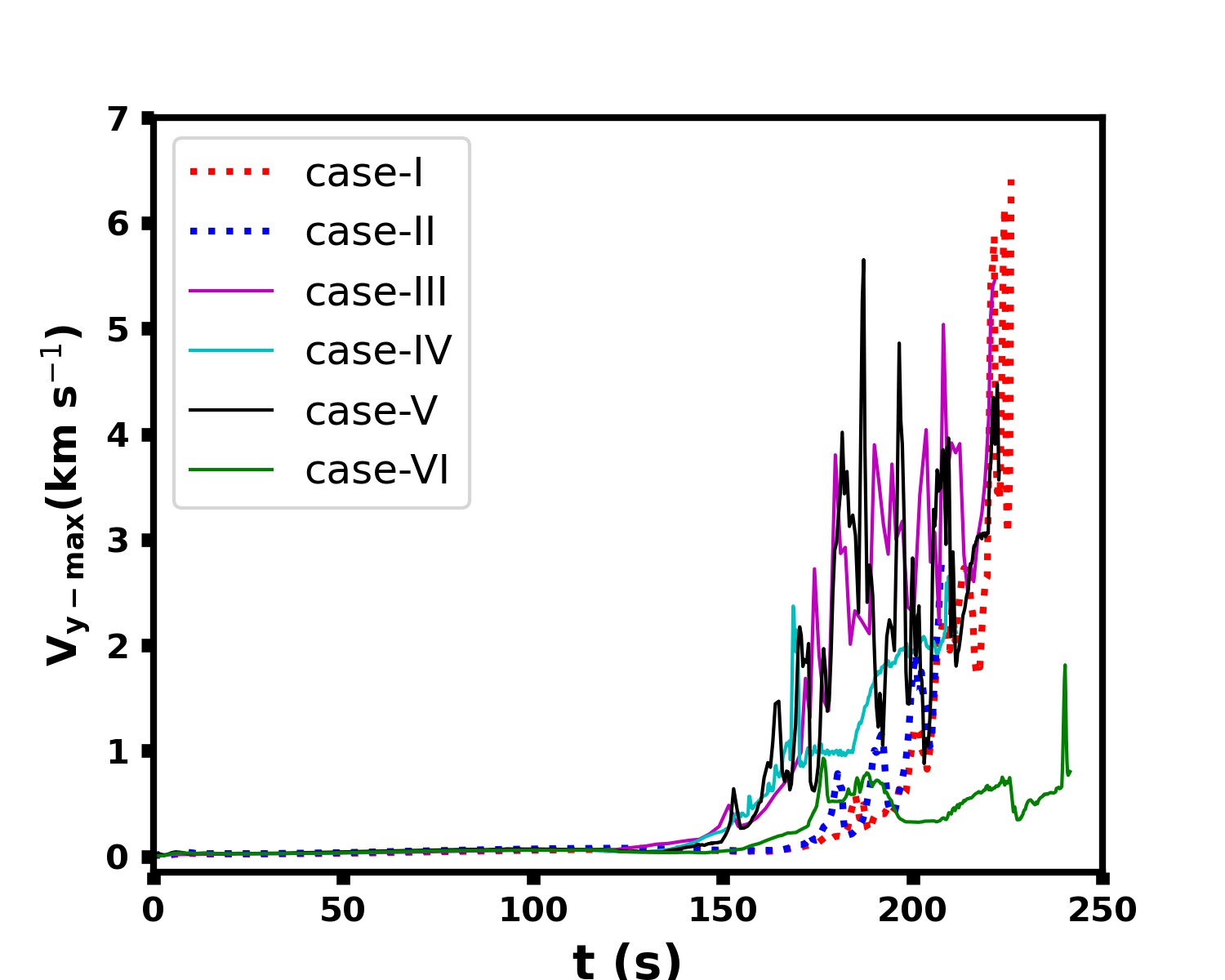}
\put(-110,145){\textbf{(c)}}
\end{minipage}
\begin{minipage}{0.495\textwidth}
\includegraphics[width=1.0\textwidth]{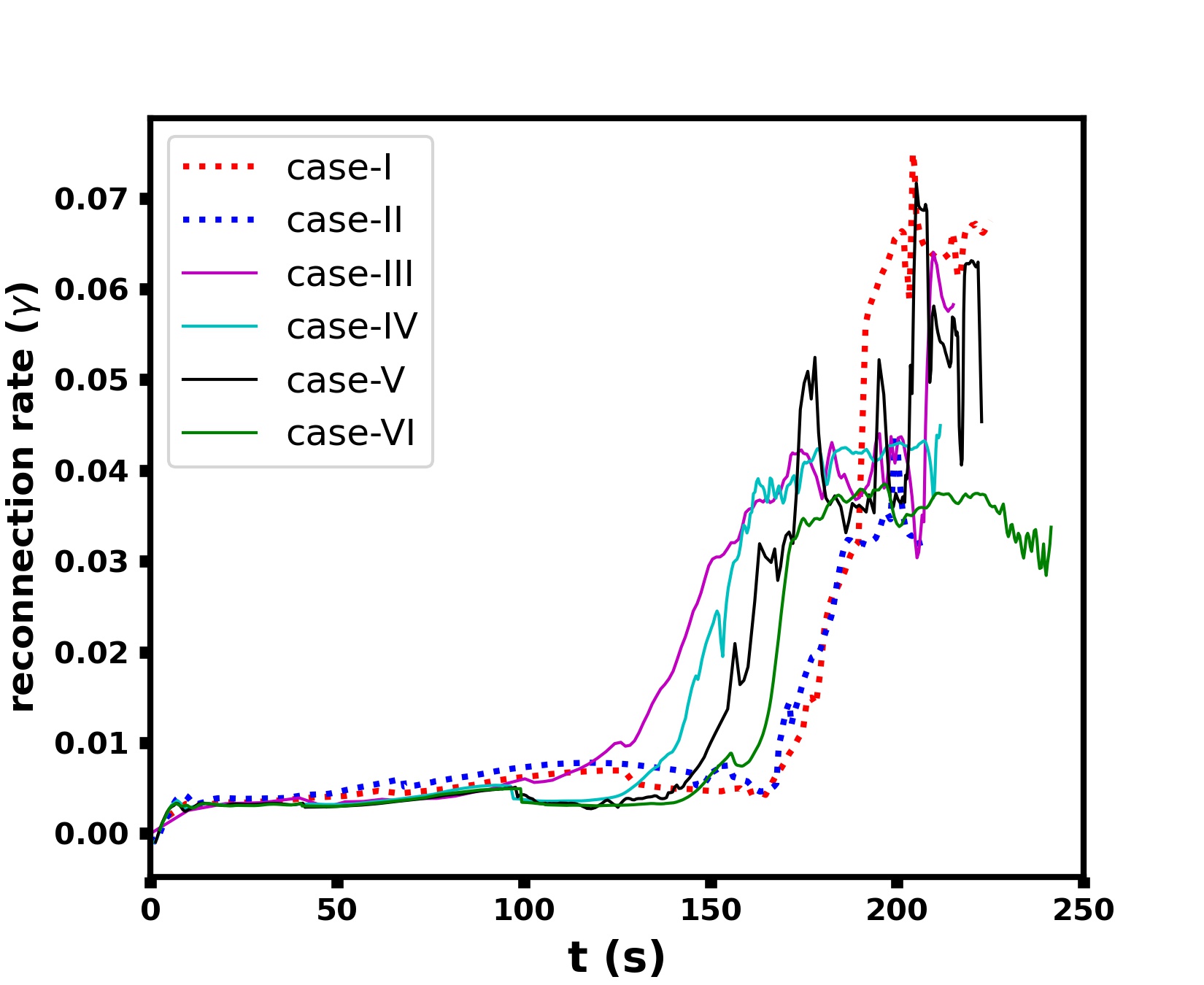}
\put(-110,145){\textbf{(d)}}
\end{minipage}
\caption{Temporal evolutions of maximum temperature $ T_{max}$ (a), maximum outflow velocity $ v_{x-max}$ (b), maximum inflow velocity $ v_{y-max}$ (c) and reconnection rate $\gamma$ (d). The cases displayed are for Z = 400 km with different cooling models (red and blue, dotted lines), Z = 800 km (magenta), Z = 1200 km (cyan), Z = 1700 km (black) and Z = 2000 km (green). The temperature increases more drastically above the middle chromosphere. Inflow velocity and reconnection rate is lowest for Z = 2000 km.}
\label{fig_2}
\end{figure}

\begin{figure}
\centering
\begin{minipage}{0.485\textwidth}
\includegraphics[width=1.0\textwidth]{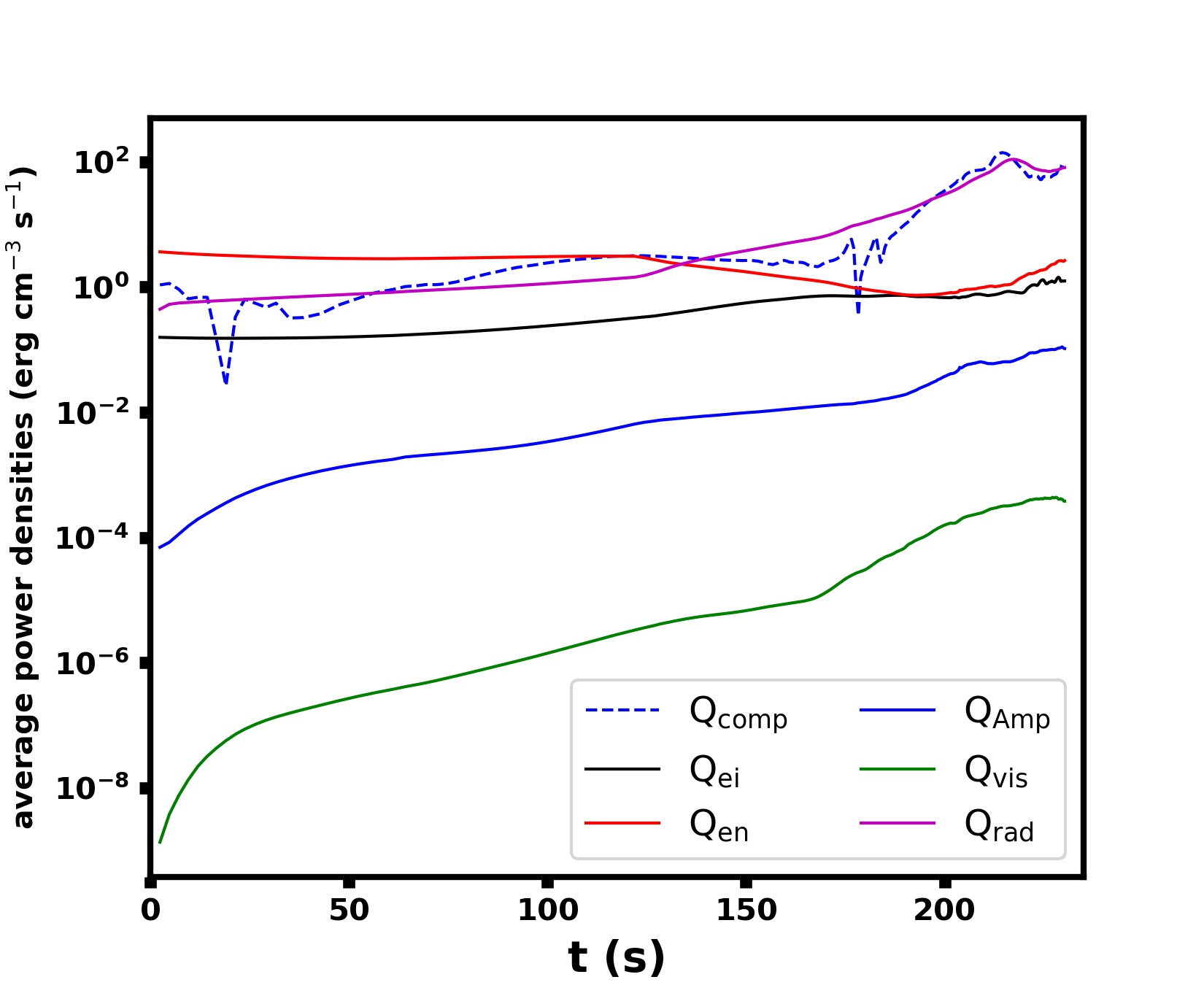}
\put(-160,145){\textbf{(a) Case-I (Z = 400 km, Q$_{rad1}$)}}
\end{minipage}
\begin{minipage}{0.485\textwidth}
\includegraphics[width=1.0\textwidth]{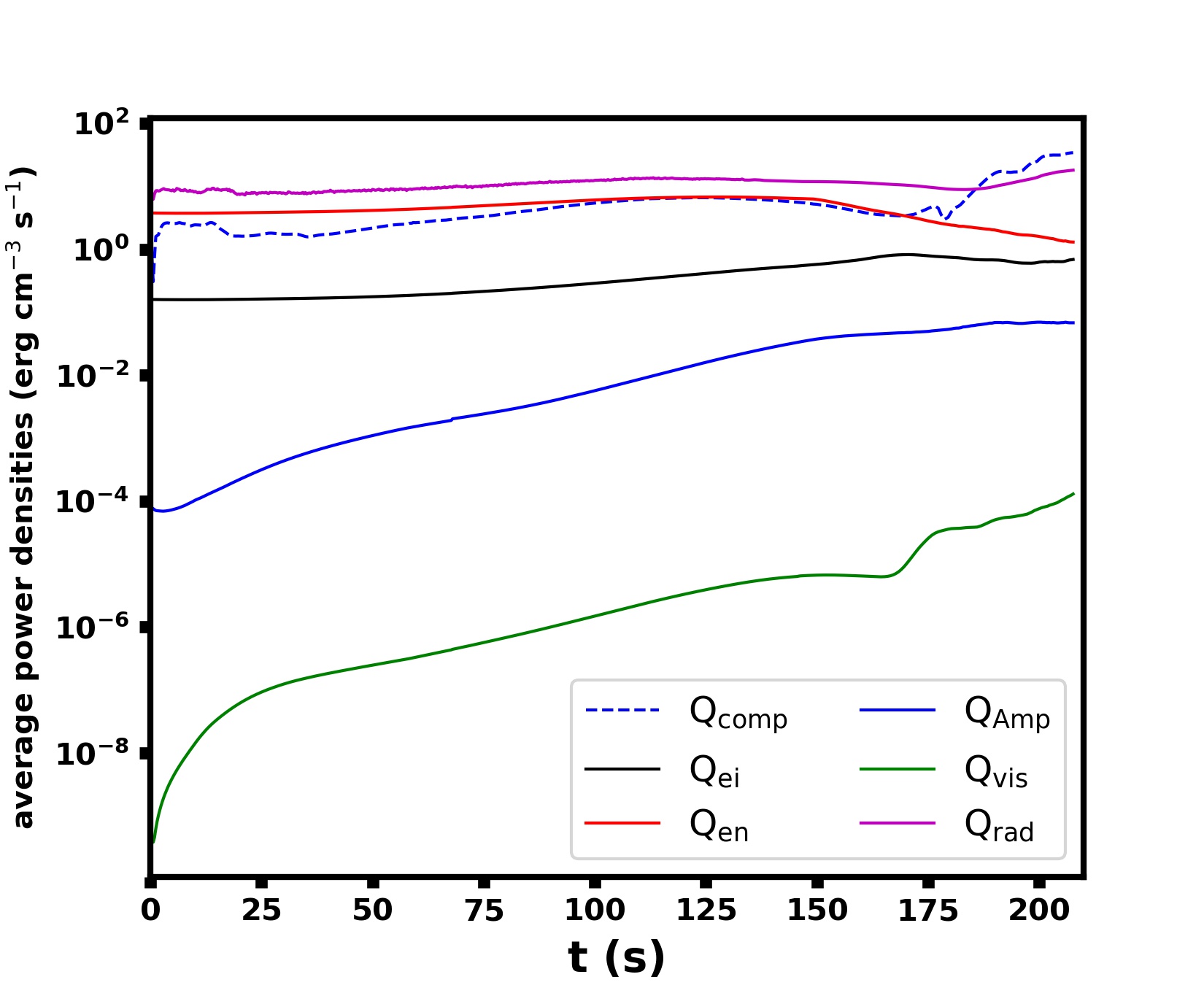}
\put(-160,145){\textbf{(b) Case-II (Z = 400 km, Q$_{rad2}$)}}
\end{minipage}
\begin{minipage}{0.485\textwidth}
\includegraphics[width=1.0\textwidth]{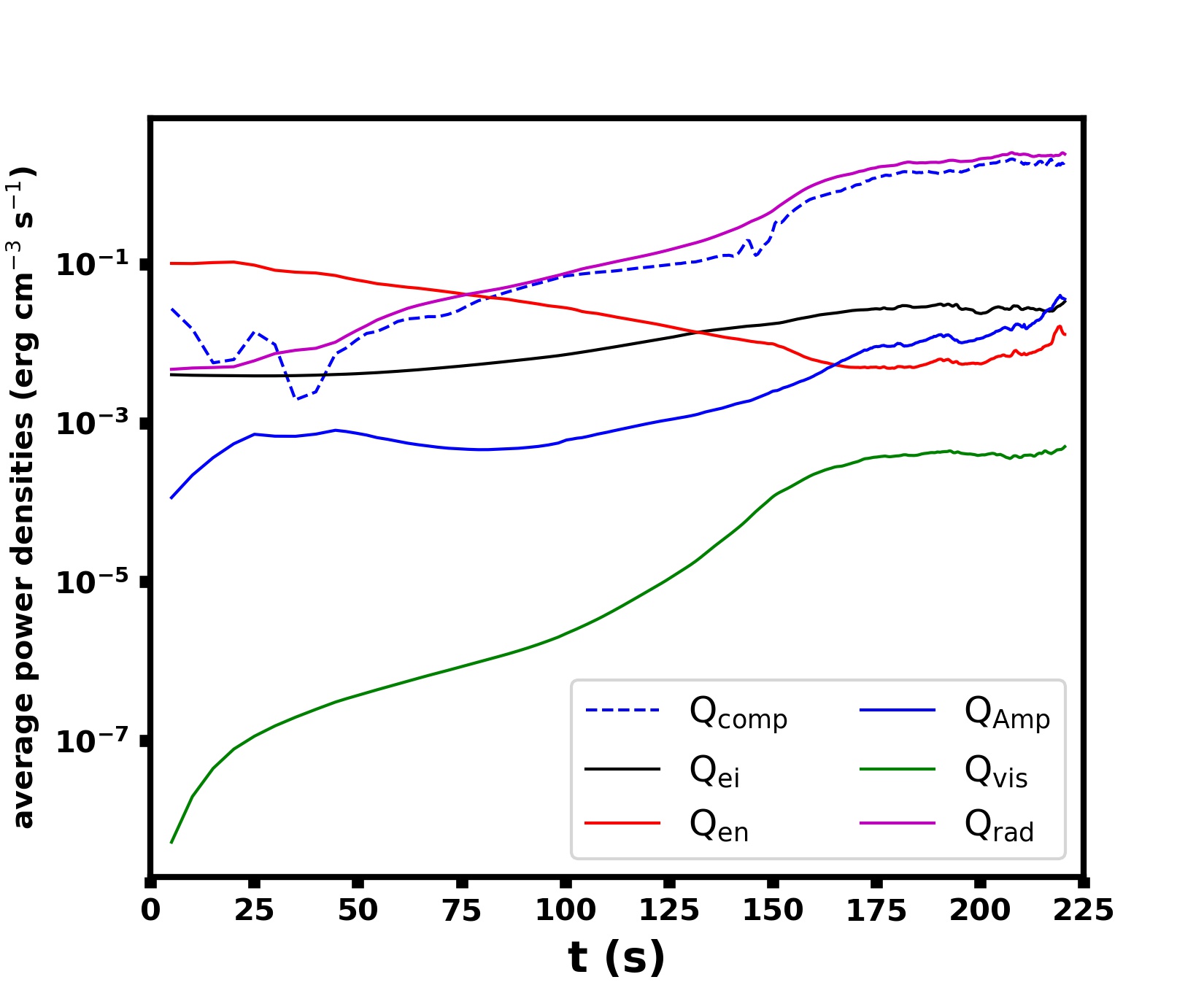}
\put(-160,145){\textbf{(c) Case-III (Z = 800 km, Q$_{rad3}$)}}
\end{minipage}
\begin{minipage}{0.485\textwidth}
\includegraphics[width=1.0\textwidth]{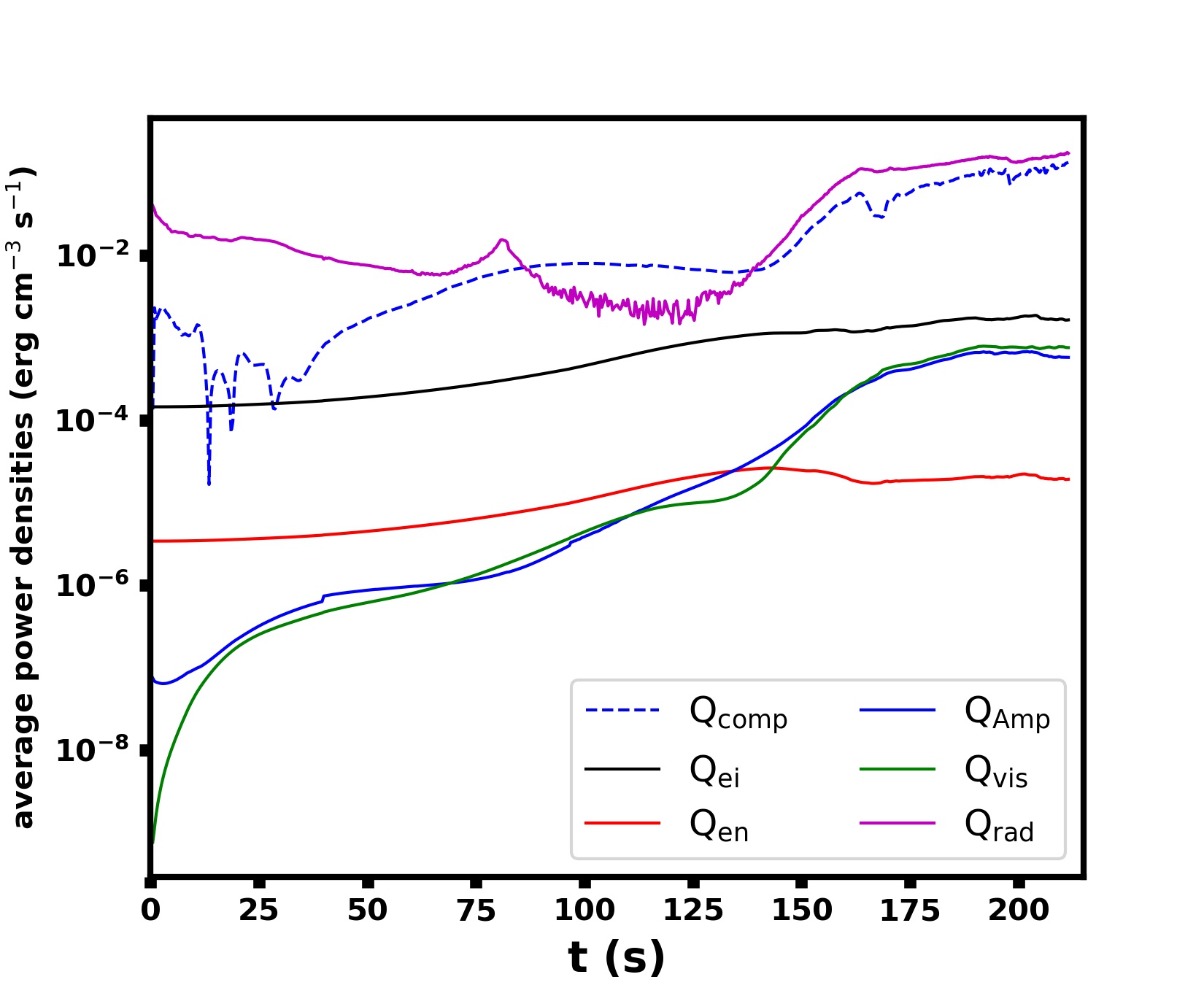}
\put(-160,145){\textbf{(d) Case-IV (Z = 1200 km, Q$_{rad3}$)}}
\end{minipage}
\begin{minipage}{0.485\textwidth}
\includegraphics[width=1.0\textwidth]{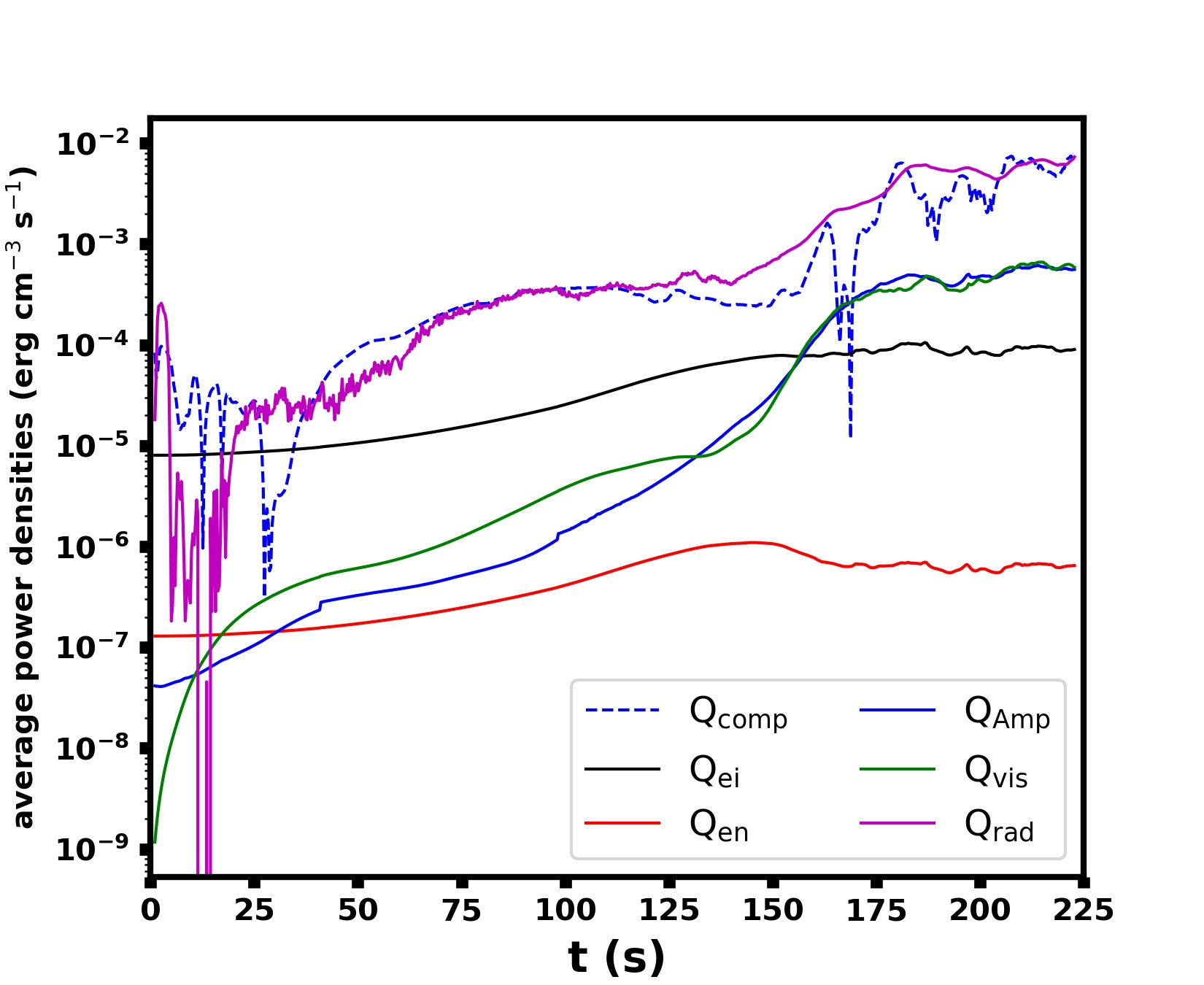}
\put(-160,145){\textbf{(e) Case-V (Z = 1700 km, Q$_{rad3}$)}}
\end{minipage}
\begin{minipage}{0.485\textwidth}
\includegraphics[width=1.0\textwidth]{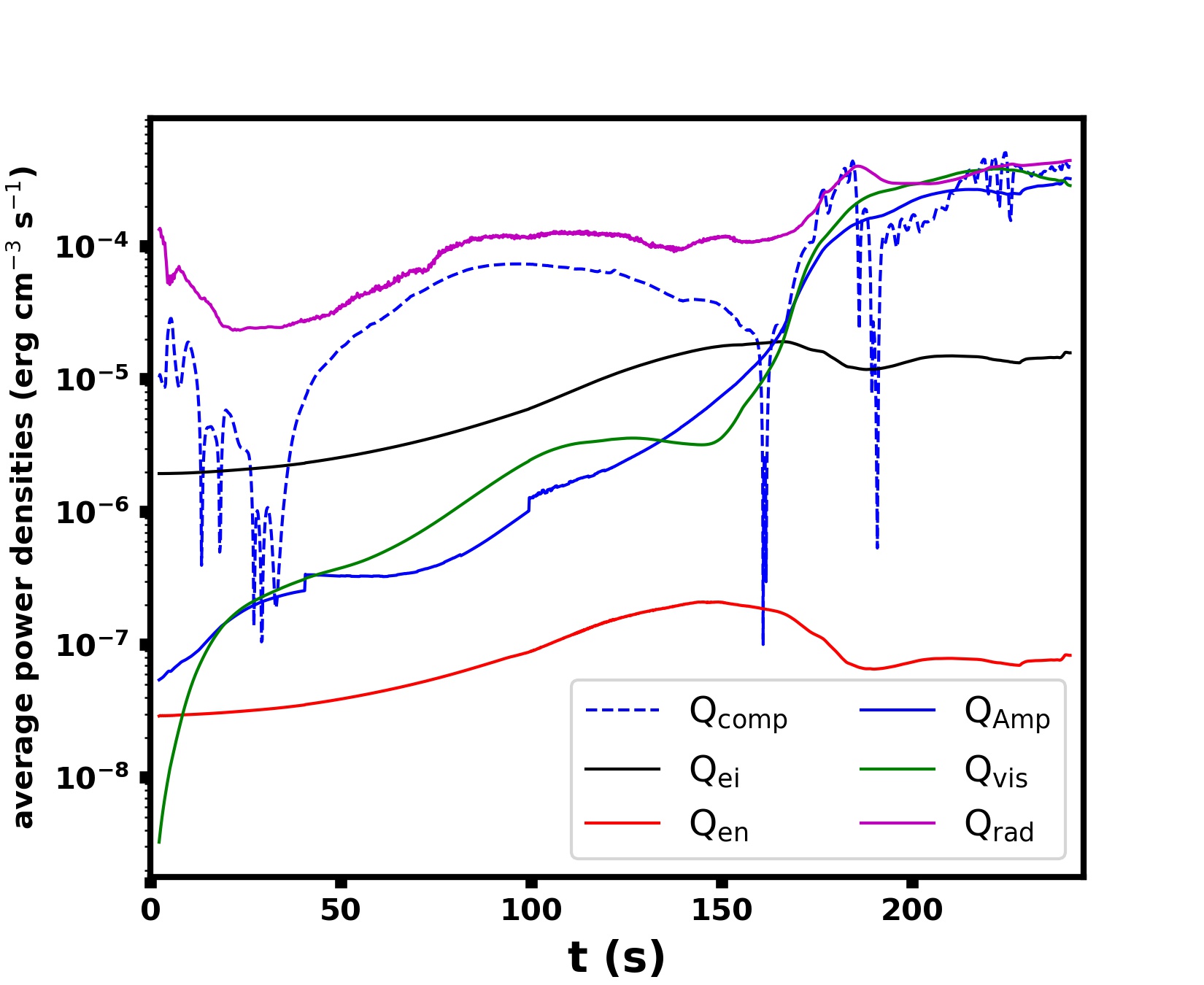}
\put(-160,145){\textbf{(f) Case-VI (Z = 2000 km, Q$_{rad3}$)}}
\end{minipage}
\caption{Temporal evolutions of the average power densities contributed by different heating terms and the radiative cooling in the photosphere at Z = 400 km above the solar surface (cases-I \& II), lower chromosphere at Z = 800 km (case-III), middle chromosphere at Z = 1200 km (case-VI) and upper chromosphere at Z = 1700 km and 2000 km (cases-V \& VI). The joule heating (Q$_{en}$) is the dominant mechanism to heat the plasma in quasi-static phase of magnetic reconnection at the photospheric and lower chromospheric heights. The compression heating (Q$_{comp}$) heats the plasma at the middle chromosphere and above through whole reconnection process. Q$_{Amp}$ and Q$_{vis}$ are equally essential in heating plasma at upper chromospheric altitudes but are insignificant at very low atmosphere.}
\label{fig_3}
\end{figure}

Temperature (upper panel) and plasma density (lower panel) distributions as well as the magnetic field topology in $xy$-plane at three different times for six cases are presented in Fig.~\ref{fig_1}.  
The current sheet becomes unstable and leads to formation of multiple magnetic islands, after the quasi-static evolution in the Sweet-Parker reconnection phase in all simulation cases. 
These islands coalesce with one another, forming bigger magnetic islands.
Plasma density and temperature distributions become non-uniform in the reconnection region after plasmoid instability occurs, much plasma is concentrated inside the plasmoids and higher temperature regions usually appear surrounding the big plasmoids. 
Comparing the temperature distributions in six cases, we find that the plasma in the reconnection region of the middle chromosphere and above (Cases IV, V and VI) is heated to a higher temperature than that in the lower chromosphere (Case III) and the photosphere (Cases I and II). 
The plasma in photospheric case using Gan \& Fang radiative cooling model (Case I) is heated to slightly higher temperature than its corresponding case with Abbet \& Fisher model (Case II) before the plasmoid instability takes place (see Figs.~\ref{fig_1}(a), \ref{fig_1}(b) and Fig.~\ref{fig_2}(a)), which is caused by the stronger radiative cooling effect in Case II during this period as shown in Figs.~\ref{fig_3}(a) and \ref{fig_3}(b).

The evolution in the maximum temperature T$_{max}$, the maximum outflow velocity V$_{x-max}$, the maximum inflow velocity V$_{y-max}$ and the magnetic reconnection rate $\gamma$ is depicted in Fig.~\ref{fig_2}.
We use the same method as Liu et al. 2023~\cite{liu2023numerical} to calculate the reconnection rate $\gamma = V_{y-aver}/V_{A-aver}$, which is defined as the ratio of the average inflow velocity ($V_{y-aver}$) to the average Alfven velocity ($V_{A-aver}$) inside a small region around the principal reconnection X-point, the red boxes in Fig.~\ref{fig_1} are these small regions for calculating the reconnection rate. All the variables shown in Fig.~\ref{fig_2} drastically increase with time after the plasmoid instability takes place.


As shown in Fig.~\ref{fig_2}(a), the temperature in the cases above the middle chromosphere increases more sharply than that in the lower chromosphere and in the photosphere after plasmoid instability occurs. 
The values of $T_{max}$ reach about 12,000 K and 14, 000 K in the later stage of the magnetic reconnection process at heights of 1700 km and 2000 km in the chromosphere, respectively. 
But the value of $T_{max}$ is always below 8000 K in such a high $\beta$ reconnection process in the lower chromosphere and in the photosphere, higher plasma density and stronger radiative cooling are the main reasons that make the temperature increase more difficult.
However, we should also note that the stronger reconnecting magnetic fields and the lower plasma $\beta$ results in the apparent temperature increase even below the middle chromosphere~\cite{ni2021magnetic}.

Figure~\ref{fig_2}(b) shows the evolution in the maximum outflow velocity, v$_{x-max}$, at the quasi-static stage until t$\sim$ 120 s is similar in all these simulations. However, the sharp rise in v$_{x-max}$ starts slightly earlier for the chromospheric cases. 
The maximum outflow velocities are almost comparable in most of the simulation cases, ranging from 8 km s$^{-1}$ to 10 km s$^{-1}$, and the values are only slightly higher above the middle chromosphere. 
The value of v$_{x-max}$ for case-II (Z = 400 km, Q$_{rad2}$) is not as high as other cases, as this simulation case terminates earlier, it is possible that v$_{x-max}$ could also reach up to 8 km s$^{-1}$ if the simulation could last longer.

The maximum inflow velocity and the reconnection rate are presented in Figs.~\ref{fig_2}(c) and \ref{fig_2}(d). 
The values of v$_{y-max}$ in the photospheric, the lower chromospheric, the middle chromospheric and the lower upper chromospheric cases (Cases I-V) can reach up to 3 km s$^{-1}$ - 6km s$^{-1}$ after plasmoid instability happens, but it is obviously small in magnetic reconnection case carried out at the top of the chromosphere, Case VI ($\sim$ 1 km s$^{-1}$). 
The maximum reconnection rate in cases I-V reaches about 0.04 - 0.07, and is also small in Case VI ($\sim$ 0.03).

The thermal energy density equation is given as below:
\begin{eqnarray}
\frac{d e_{th}}{dt} = -p \nabla \cdot \mathbf{v} + \frac{1}{2 \xi} Tr(\tau^{2}_{S}) + \frac{\eta}{\mu_{0}} |\nabla \times \mathbf{B}|^{2} + \frac{\eta_{AD}}{\mu_{0}^{2}} |\mathbf{B} \times (\nabla \times \mathbf{B})|^{2} + Q_{rad}. 
\label{eq:33}
\end{eqnarray}
Here $e_{th}$ represents the thermal energy density and $T_{r}$ denotes the trace of stress tensor. 
The terms on the right hand side of the above equation contains various terms of thermal energy gain such as, compression heating, Q$_{comp}$ = $-p \nabla \cdot \mathbf{v}$, viscous heating, Q$_{vis}$ = $\frac{1}{2 \xi} Tr(\tau^{2}_{S})$, ohmic (joule) heating contributed by magnetic diffusions $\eta_{ei}$ and $\eta_{en}$, Q$_{ei,en}$ =  $\frac{\eta_{ei,en}}{\mu_{0}} |\nabla \times \mathbf{B}|^{2}$ as well as frictional heating contributed by the ambipolar diffusion, Q$_{Amp}$ =  $\frac{\eta_{AD}}{\mu_{0}^{2}} |\mathbf{B} \times (\nabla \times \mathbf{B})|^{2}$. The final term $Q_{rad}$ describes the effect of radiative cooling. 
 The average power densities of different heating and cooling terms for six cases presented in Fig.~\ref{fig_3} are calculated by using the same method as described by Ni et. al~\cite{ni2022plausibility}. 
Here, the average power densities are calculated inside the main reconnection region within the simulation box for $0 \leq x \leq L_{0}$ and $-0.05L_{0} \leq y \leq 0.05L_{0}$.

\begin{figure}
\centering
\begin{minipage}{0.49\textwidth}
\includegraphics[width=1.0\textwidth]{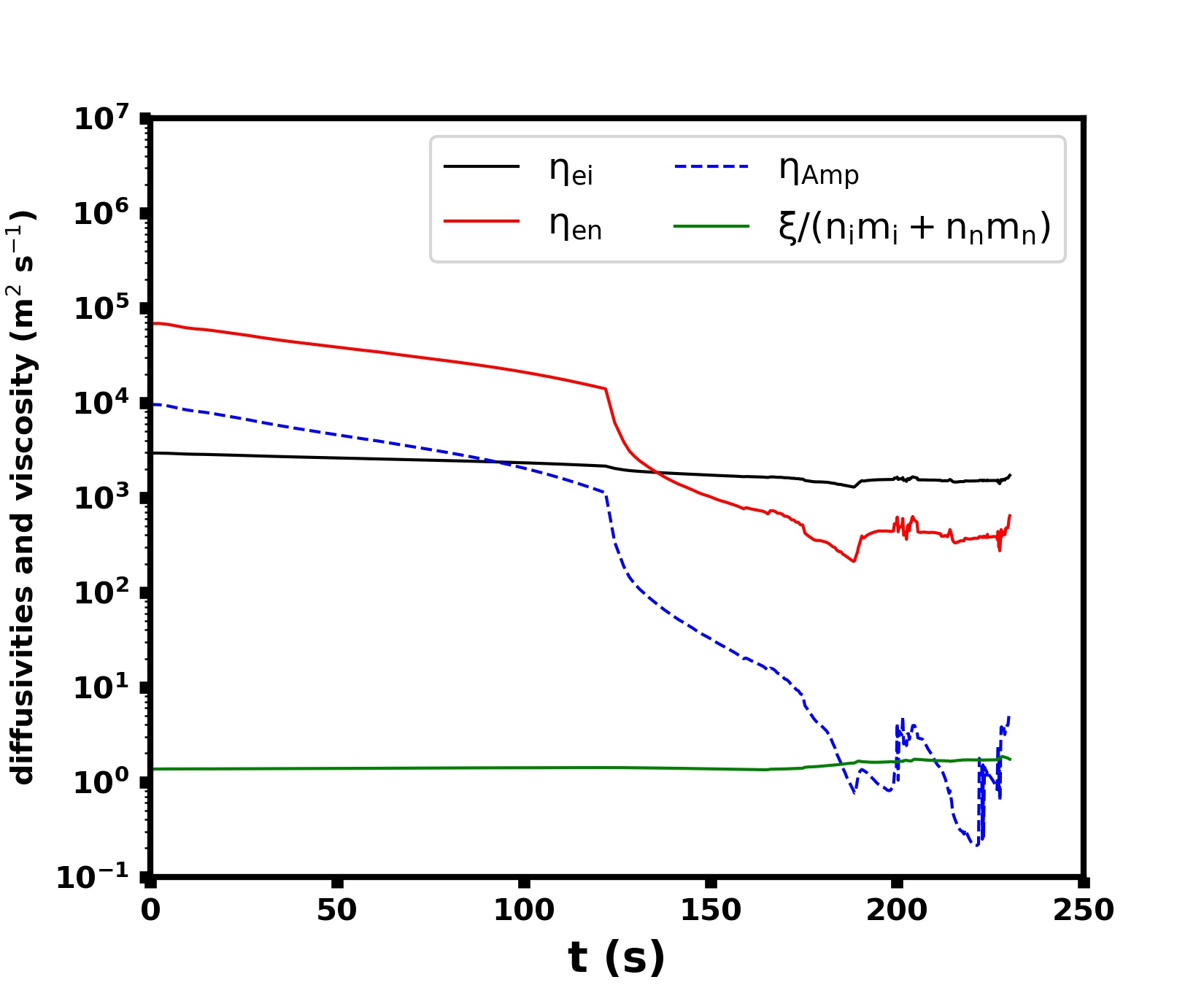}
\put(-160,145){\textbf{(a) Case-I (Z = 400 km, Q$_{rad1}$)}}
\end{minipage}
\begin{minipage}{0.49\textwidth}
\includegraphics[width=1.0\textwidth]{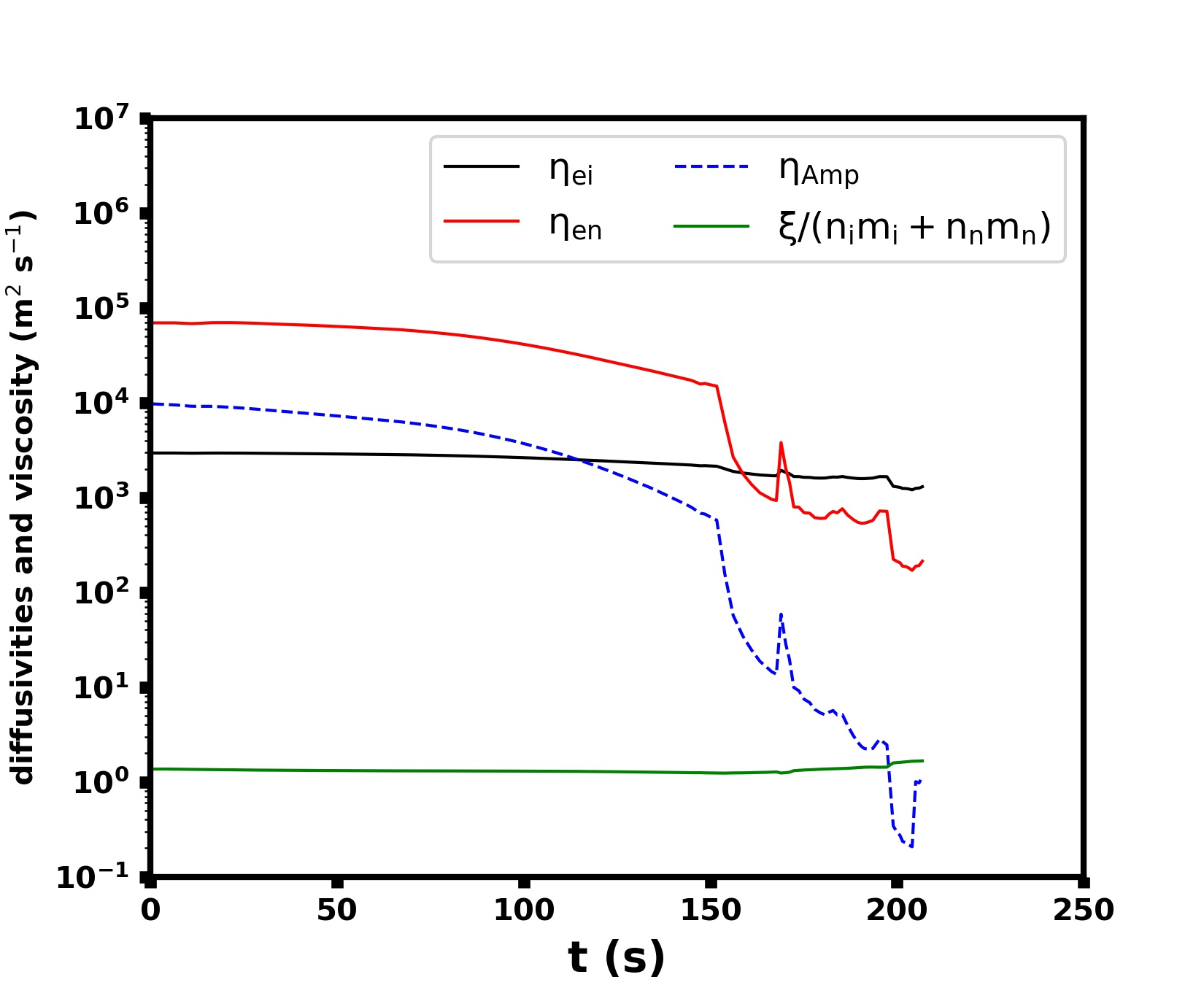}
\put(-160,145){\textbf{(b) Case-II (Z = 400 km, Q$_{rad2}$)}}
\end{minipage}
\begin{minipage}{0.49\textwidth}
\includegraphics[width=1.0\textwidth]{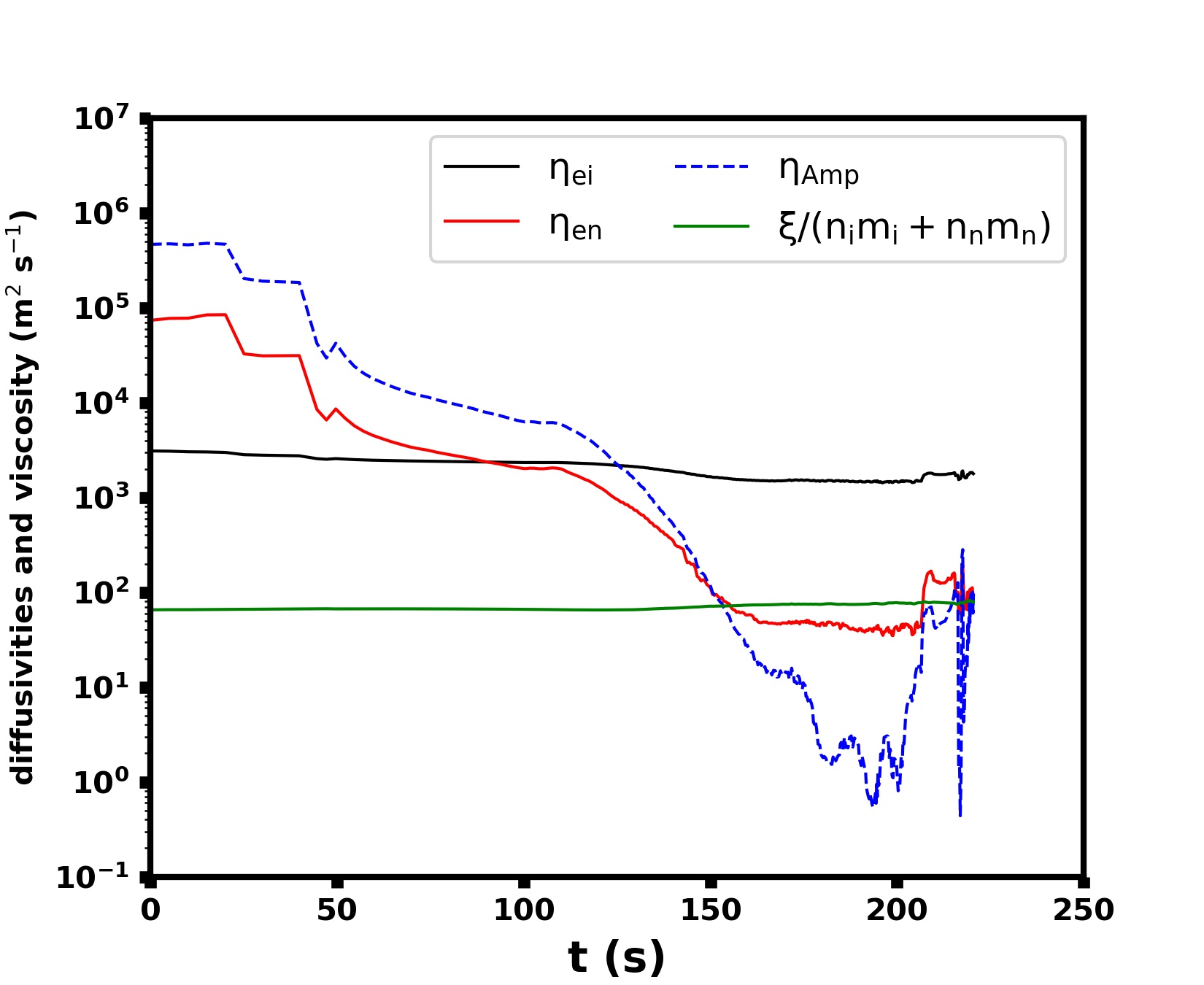}
\put(-160,145){\textbf{(c) Case-III (Z = 800 km, Q$_{rad3}$)}}
\end{minipage}
\begin{minipage}{0.49\textwidth}
\includegraphics[width=1.0\textwidth]{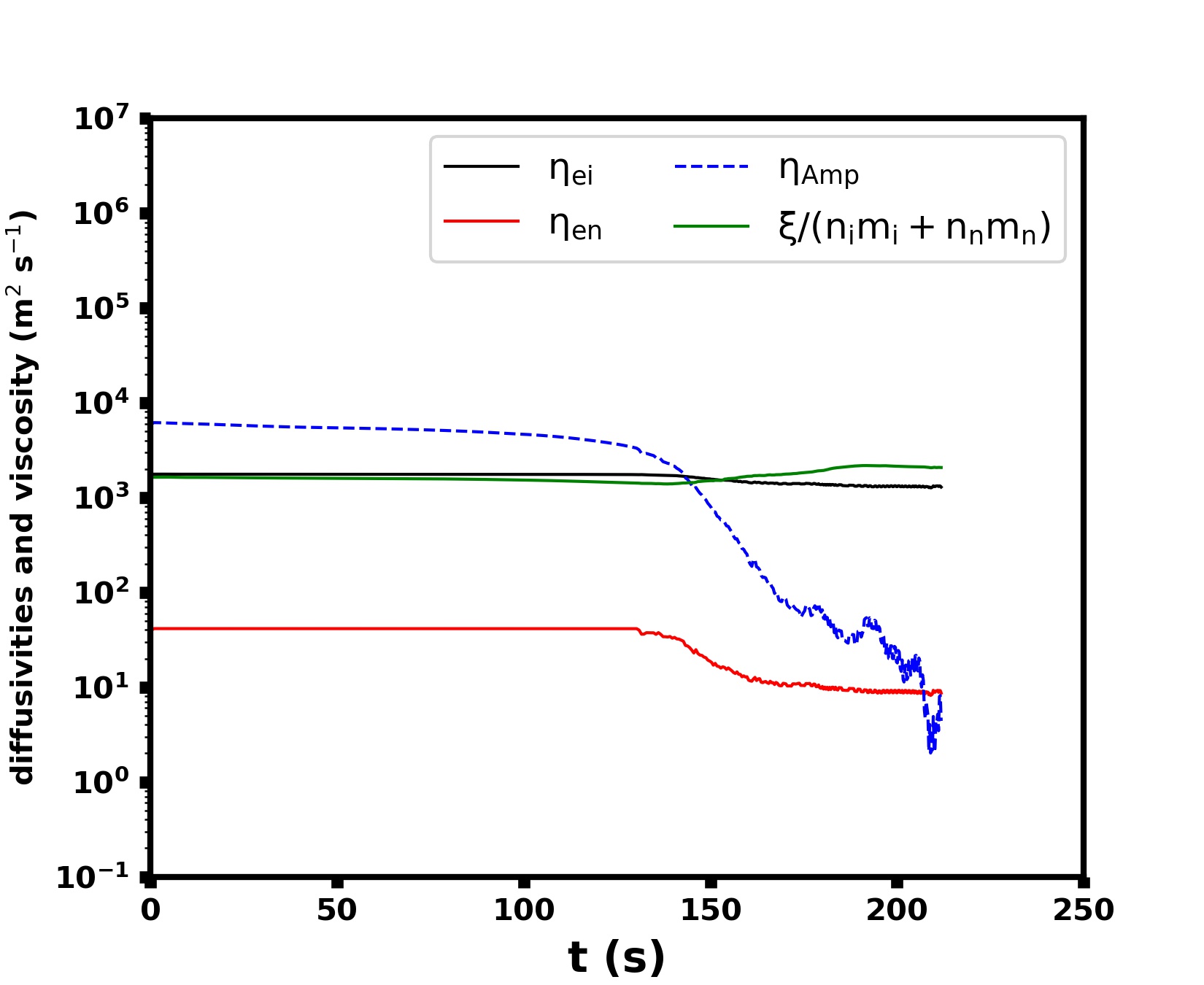}
\put(-160,145){\textbf{(d) Case-IV (Z = 1200 km, Q$_{rad3}$)}}
\end{minipage}
\begin{minipage}{0.49\textwidth}
\includegraphics[width=1.0\textwidth]{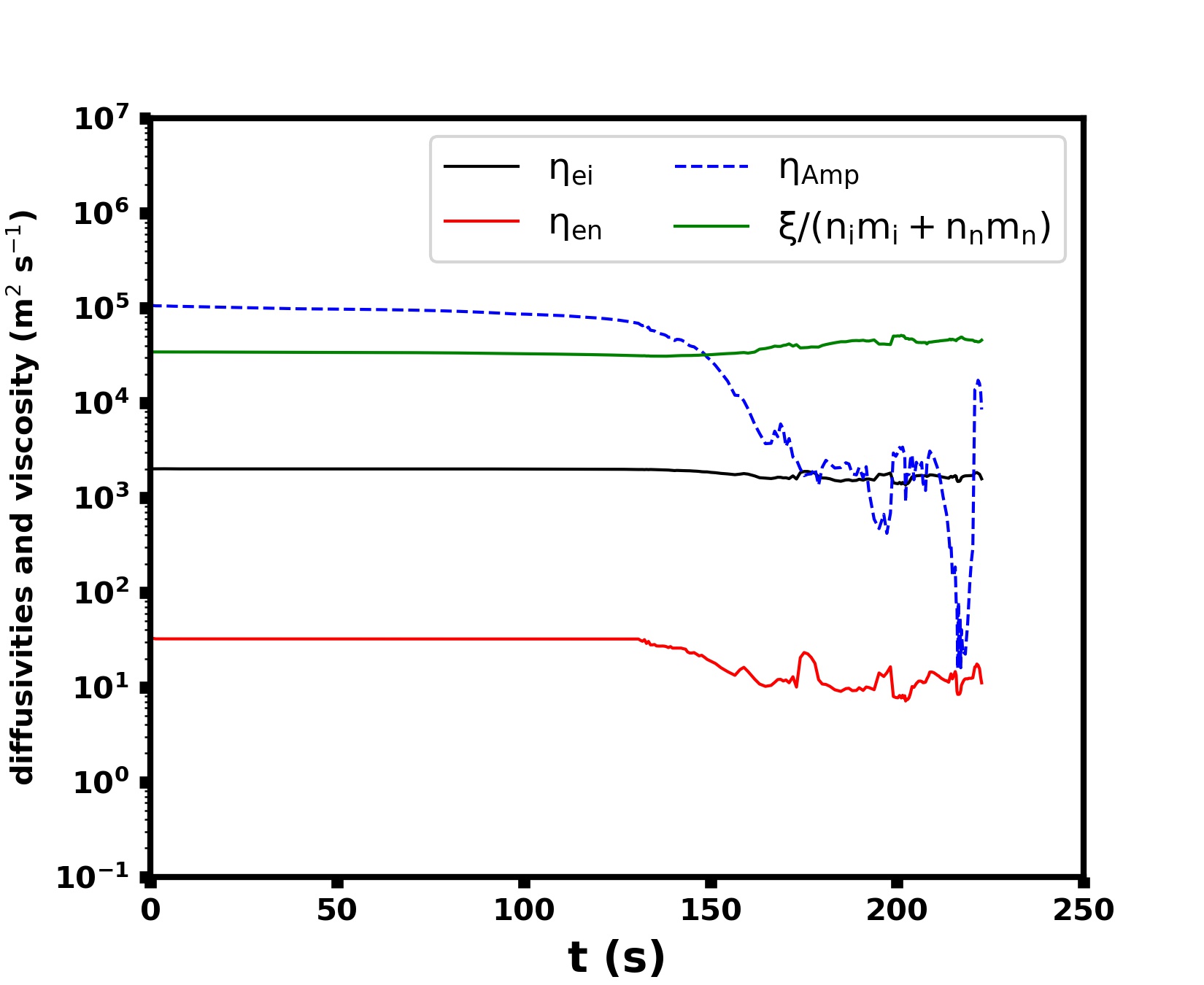}
\put(-160,145){\textbf{(e) Case-V (Z = 1700 km, Q$_{rad3}$)}}
\end{minipage}
\begin{minipage}{0.49\textwidth}
\includegraphics[width=1.0\textwidth]{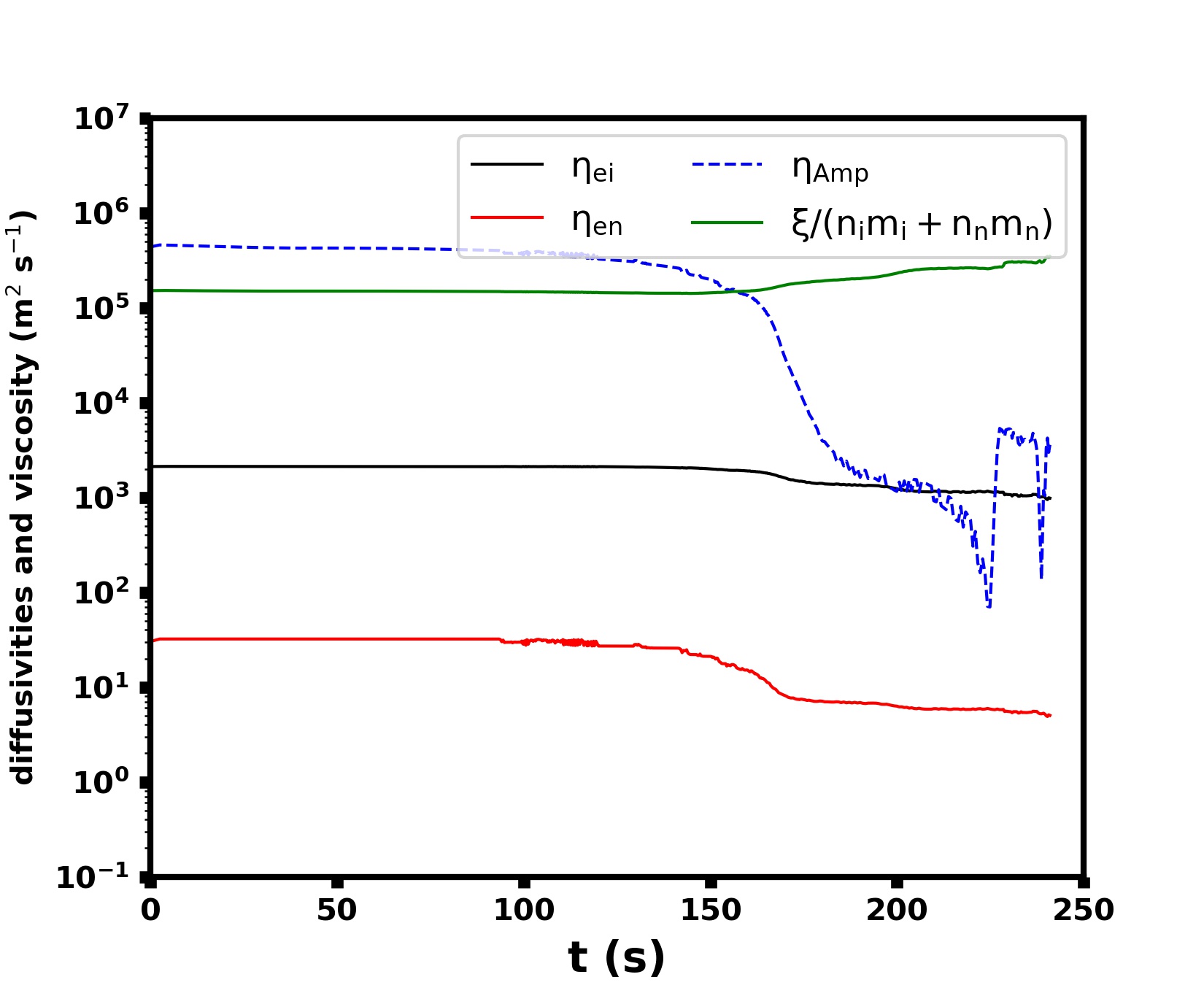}
\put(-160,145){\textbf{(f) Case-VI (Z = 2000 km, Q$_{rad3}$)}}
\end{minipage}
\caption{Temporal evolutions of different diffusion coefficients ($\eta_{ei}$, $\eta_{en}$, $\eta_{AP}$) and the coefficient related to viscosity ($\xi/(n_{i}m_{i}+n_{n}m_{n})$) at the main reconnection X-point in the photosphere at Z = 400 km above the solar surface (cases-I \& II), lower chromosphere at Z = 800 km (case-III), middle chromosphere at Z = 1200 km (case-VI) and upper chromosphere at Z = 1700 km and 2000 km (cases-V \& VI). The ambipolar diffusion ($\eta_{Amp}$) is larger than other diffusion terms in the beginning in all chromospheric simulations. The effect of viscosity increases as the reconnection region movers to the higher heights.}
\label{fig_4}
\end{figure}


Figures~\ref{fig_3}(a)-\ref{fig_3}(c) exhibit that Q$_{en}$ is larger than the other heating terms in the beginning of the magnetic reconnection process. 
This means that the Joule heating as a result of the electron-neutral collision dominates initially and play an important role in heating the plasma in the quasi-static stage of magnetic reconnection at the photosphere and at the bottom of the chromosphere. 
In the later stage, Q$_{en}$ decreases while Q$_{comp}$ increases and becomes a primary heating contributor for the rest of the reconnection process.
Small values of Q$_{Amp}$ and Q$_{vis}$ during the reconnection process indicate that their contributions to heating plasma at photospheric and lower chromospheric altitudes are not apparent. 
These results are consistent with the recent studies about magnetic reconnection in EBs~\cite{liu2023numerical}.


In contrast, the evolution of average power densities behaves very differently in the middle and upper chromosphere. 
Q$_{comp}$ becomes the dominant heating mechanism to heat plasma throughout the reconnection process (Figs.~\ref{fig_3}(d)-\ref{fig_3}(f)).    
The impact of Q$_{Amp}$ and Q$_{vis}$ is not very important in heating the photospheric and lower/middle chromospheric plasmas, their contributions are much smaller than that of Q$_{comp}$. 
However, the evolution in ambipolar and viscous heatings depicted in Figs.~\ref{fig_3}(e) and \ref{fig_3}(f) signifies that these terms are equally important for plasma heating in the upper chromosphere.
After plasmoid instability occurs, the contribution of $Q_{Amp}$ and $Q_{vis}$ are approximately equal to that of $Q_{comp}$ for Case VI (Z = 2000 km) during the later stage.

In partially ionized plasma, interactions between charged and neutral species are of key interest, it may have impact on the reconnection process. 
The evolution in diffusivities caused by electron-ion collision ($\eta_{ei}$), electron-neutral collision ($\eta_{en}$) and the ambipolar diffusion ($\eta_{Amp} = \eta_{AD}B^{2}/\mu_{0}$) caused by the decoupling of ions and neutrals at the principal reconnection X-point for six different simulation cases are displayed in Fig.~\ref{fig_4}.
All these diffusivities at the principal X-point decrease with time because of the increasing temperature.
In both photospheric reconnection cases (cases I and II), $\eta_{en}$ is initially the dominant diffusivity and is about one order of magnitude greater than $\eta_{Amp}$, while $\eta_{Amp}$ is larger than $\eta_{ei}$ (Figs.~\ref{fig_4}(a) and \ref{fig_4}(b)). 
The population of neutral particles decreases as the ionization degree of helium and hydrogen rises with temperature.
Therefore, diffusions due to the electron-neutral collision ($\eta_{en}$) and the decoupling of ions and neutrals $\eta_{Amp}$ fall off sharply in the later stage.
It is evident from Figs.~\ref{fig_4}(a) and \ref{fig_4}(b), that the evolution pattern of the diffusivities are similar in photospheric cases. 
It is worth noting that $\eta_{en}$ and $\eta_{Amp}$ drops earlier in Case I (Z = 400 km, Q$_{rad1}$) because the plasmoid instability in this case developed slightly faster. 
Figures~\ref{fig_4}(c)-\ref{fig_4}(f) show that at the beginning, the ambipolar diffusion term $\eta_{Amp}$ is dominant in all chromospheric cases. 
The value of $\eta_{en}$ at the middle and upper chromospheric heights (cases IV-VI) is much smaller than those of $\eta_{Amp}$ and $\eta_{ei}$ throughout the reconnection process. 
Whereas, the value of $\eta_{en}$ at the beginning is larger than that of $\eta_{ei}$ at the low chromospheric height (Case III).

The time-dependent coefficient relating to the viscosity $[\xi/(n_{i}m_{i}+n_{n}m_{n})]$ at the principal X-point for the six cases is also presented in Fig.~\ref{fig_4}. 
One can find that this coefficient is ignorable in the photosphere, but it reaches large values in the upper chromosphere, which makes the magnetic Prandtl number $P_{r} = \xi/[(n_{i}m_{i}+n_{n}m_{n})\eta]$ there to be large compared to unity. 
The effect of viscosity might have strong effects on magnetic reconnection and heating process in this region, which will be discussed later.

\section{Discussions}
\label{sec_IV}
\subsection{Fast reconnection mechanisms at different layers}
The mechanism that leads to fast magnetic reconnection to explain the burst like events in the universe is always one of the key issues. 
The fast reconnection mechanism in the dense and cool solar atmosphere could be very different from the solar corona. 
We need to consider the effects of radiative cooling and neutral particles on magnetic reconnection process in this region, the varying of plasma parameters with altitude might also result in varying reconnection mechanisms at different layers.

As shown in Figs.~\ref{fig_2}(c) and ~\ref{fig_2}(d), the inflow velocities and the reconnection rates at different altitudes all sharply increased to high values after the plasmoid instability takes place. 
These results demonstrate that the plasmoid instability plays a vital role in accelerating magnetic reconnection in the partially ionized plasma in the low solar atmosphere, which is consistent with previous single-fluid~\cite{ni2015fast} and two-fluids~\cite{ni2018onset} MHD simulation results in the low $\beta$ environment around the solar TMR. However, previous two-fluids MHD simulations with high plasma $\beta$ showed that the ion recombination effect also leads to fast magnetic reconnection~\cite{leake2012multi,leake2013magnetic}. 
More detailed two-fluid MHD simulations with suitable radiative cooling models are needed to check which effect is more important in such a high $\beta$ case. We also note that the inflow velocity and the reconnection rates are smaller at the high altitude in the upper chromosphere.
\begin{figure}
\centering
\begin{minipage}{0.49\textwidth}
\includegraphics[width=1.0\textwidth]{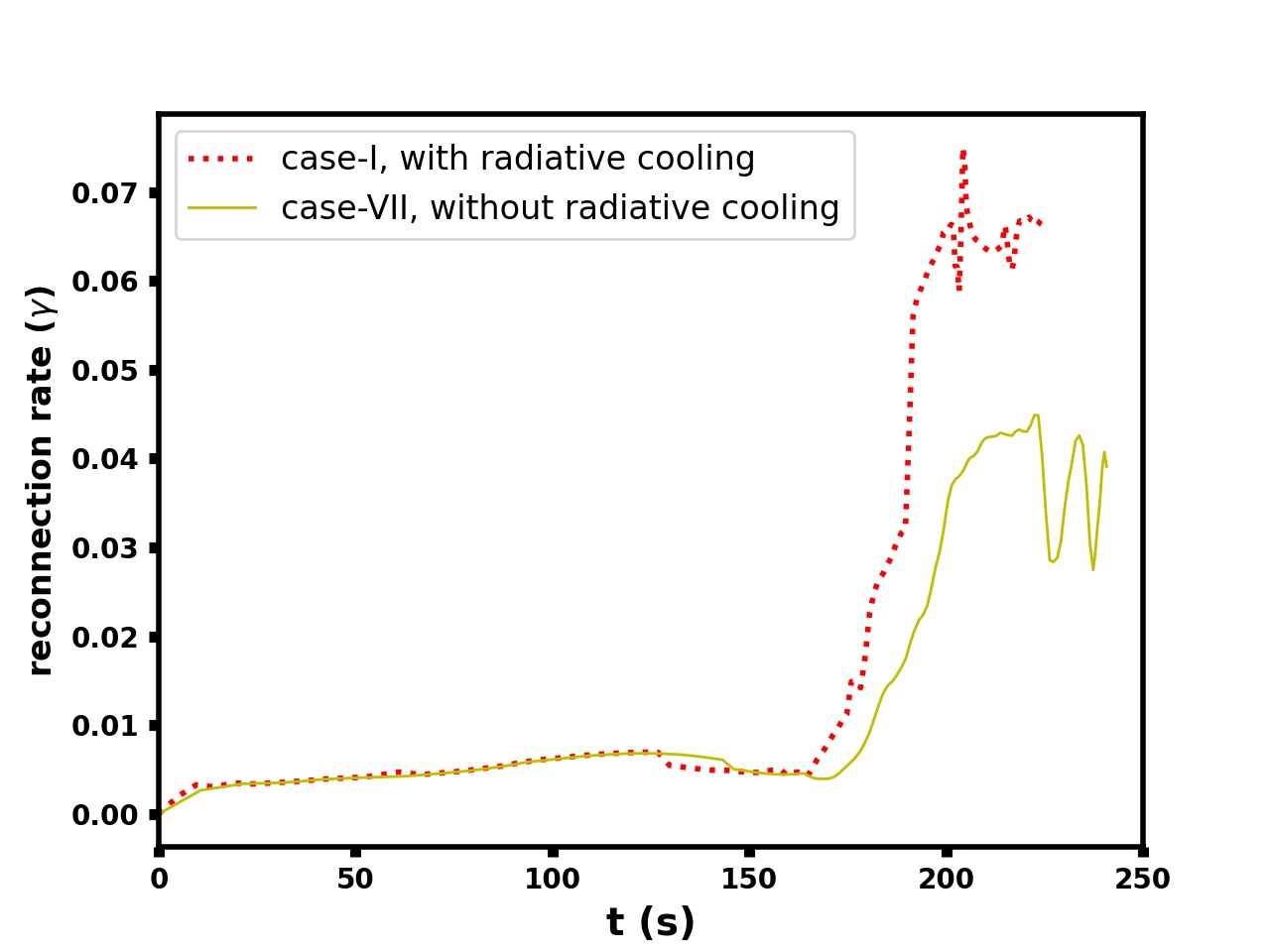}
\put(-130,135){\textbf{(a) Z = 400 km}}
\end{minipage}
\begin{minipage}{0.49\textwidth}
\includegraphics[width=1.0\textwidth]{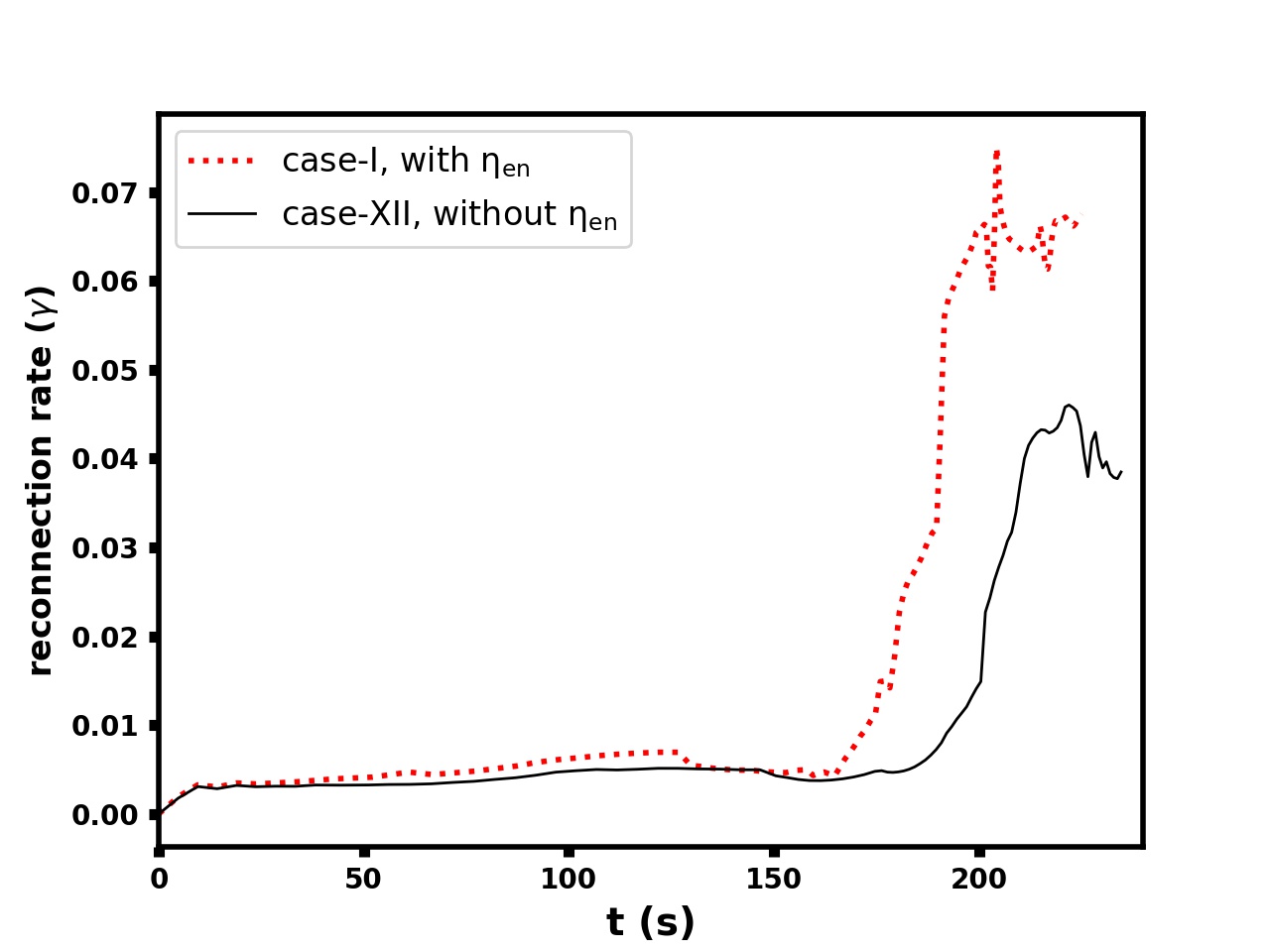}
\put(-130,135){\textbf{(b) Z = 400 km}}
\end{minipage}
\begin{minipage}{0.49\textwidth}
\includegraphics[width=1.0\textwidth]{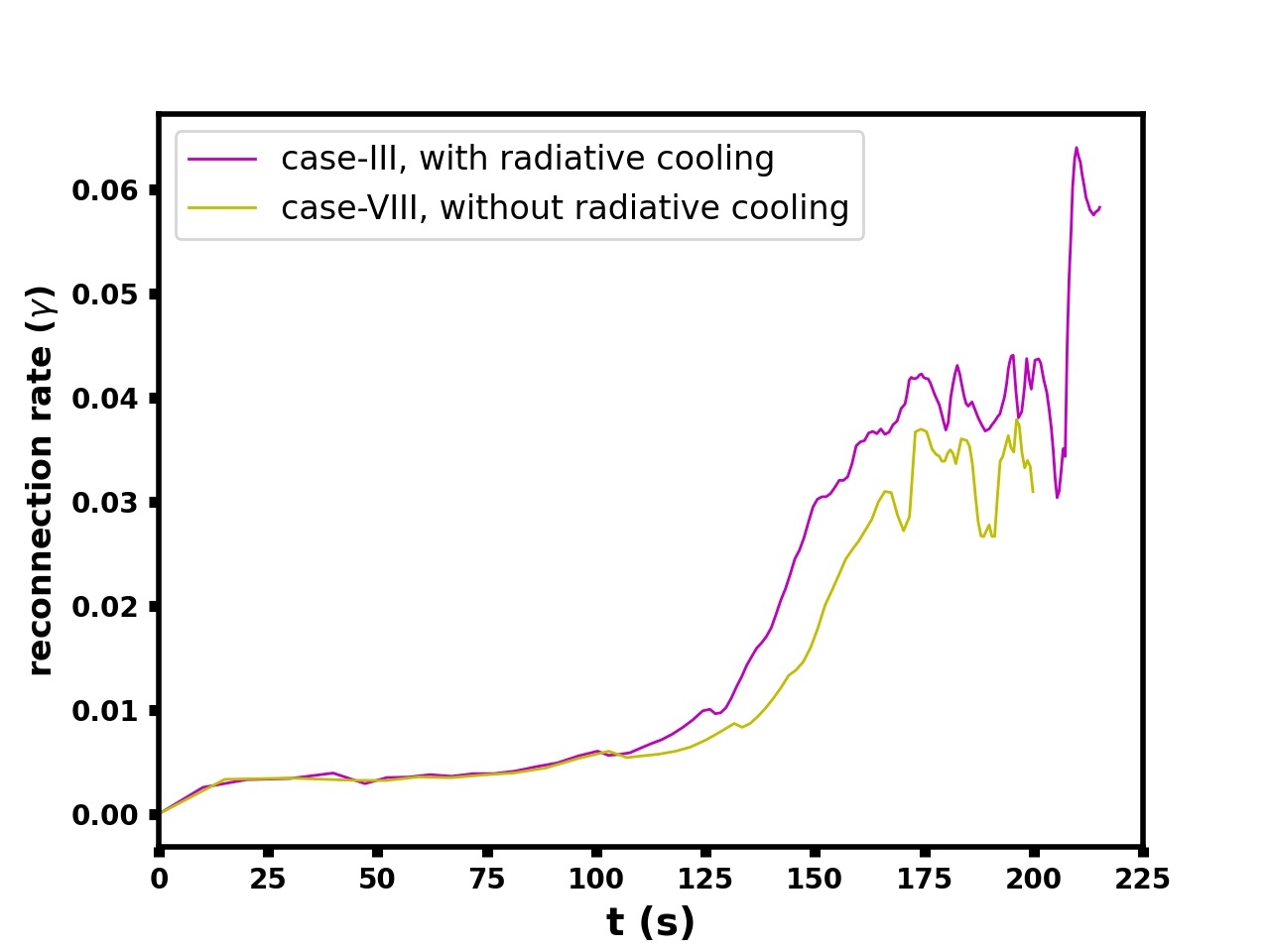}
\put(-130,135){\textbf{(c) Z = 800 km}}
\end{minipage}
\begin{minipage}{0.49\textwidth}
\includegraphics[width=1.0\textwidth]{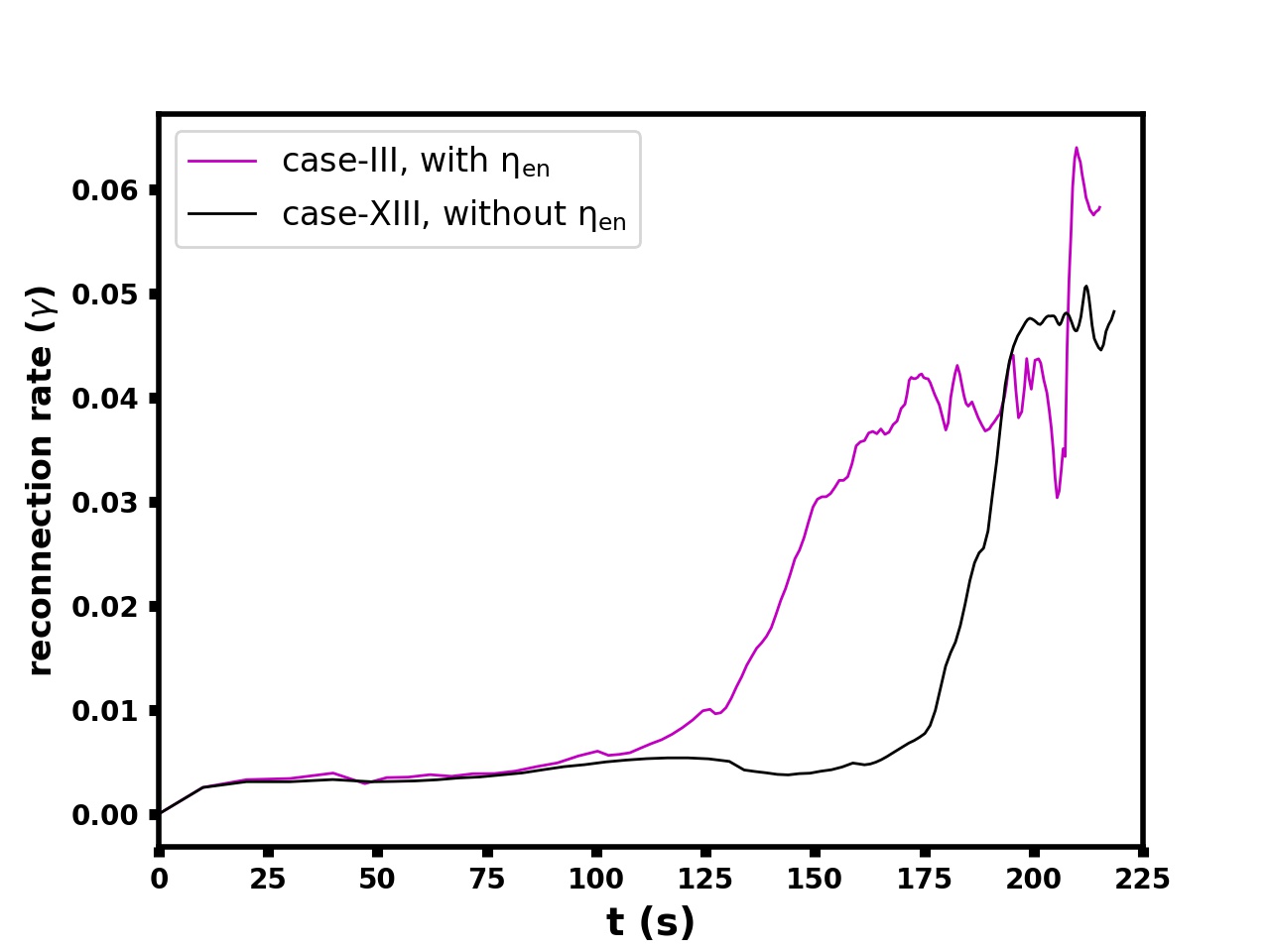}
\put(-130,135){\textbf{(d) Z = 800 km}}
\end{minipage}
\begin{minipage}{0.49\textwidth}
\includegraphics[width=1.0\textwidth]{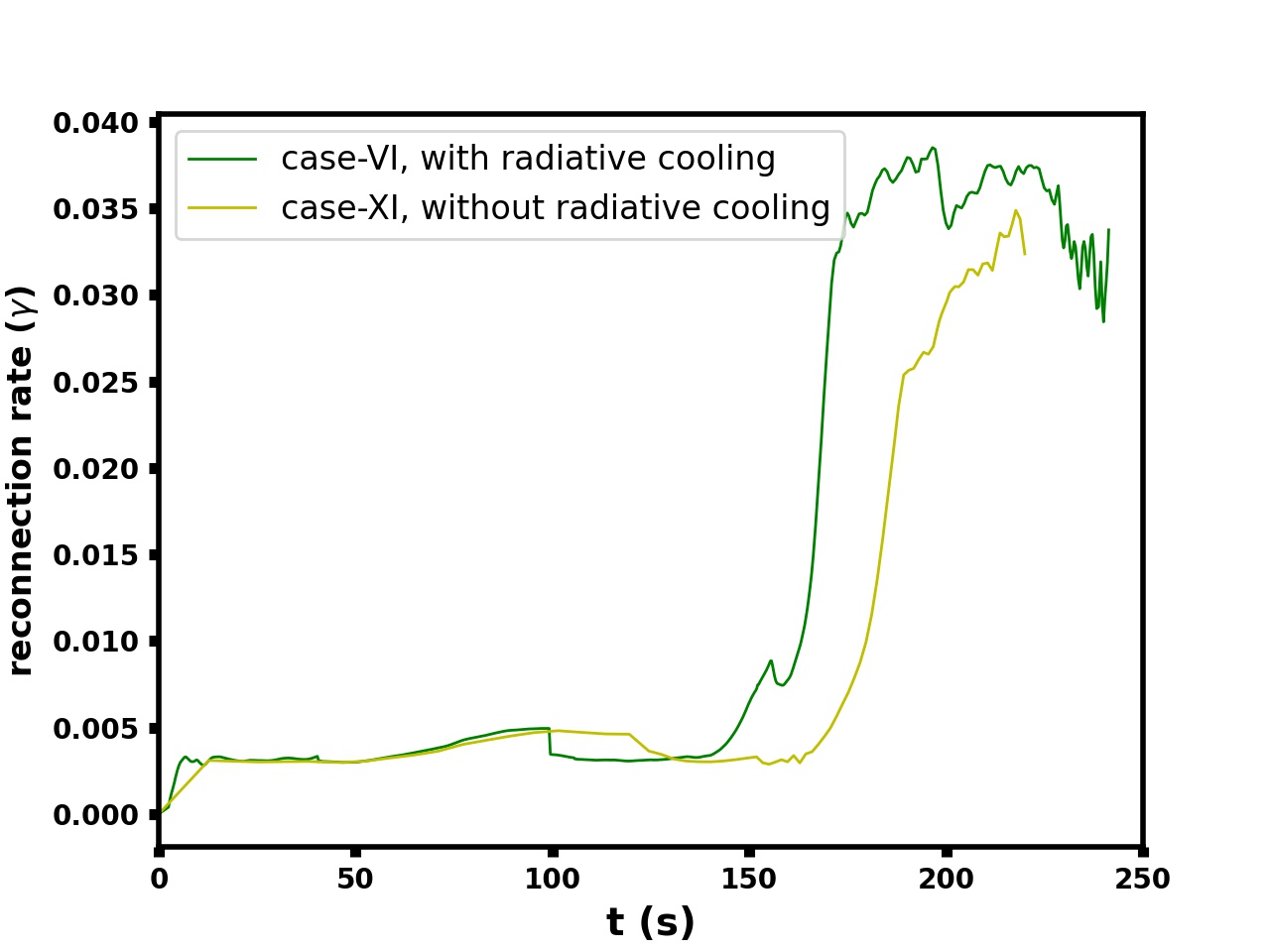}
\put(-130,135){\textbf{(e) Z = 2000 km}}
\end{minipage}
\begin{minipage}{0.49\textwidth}
\includegraphics[width=1.0\textwidth]{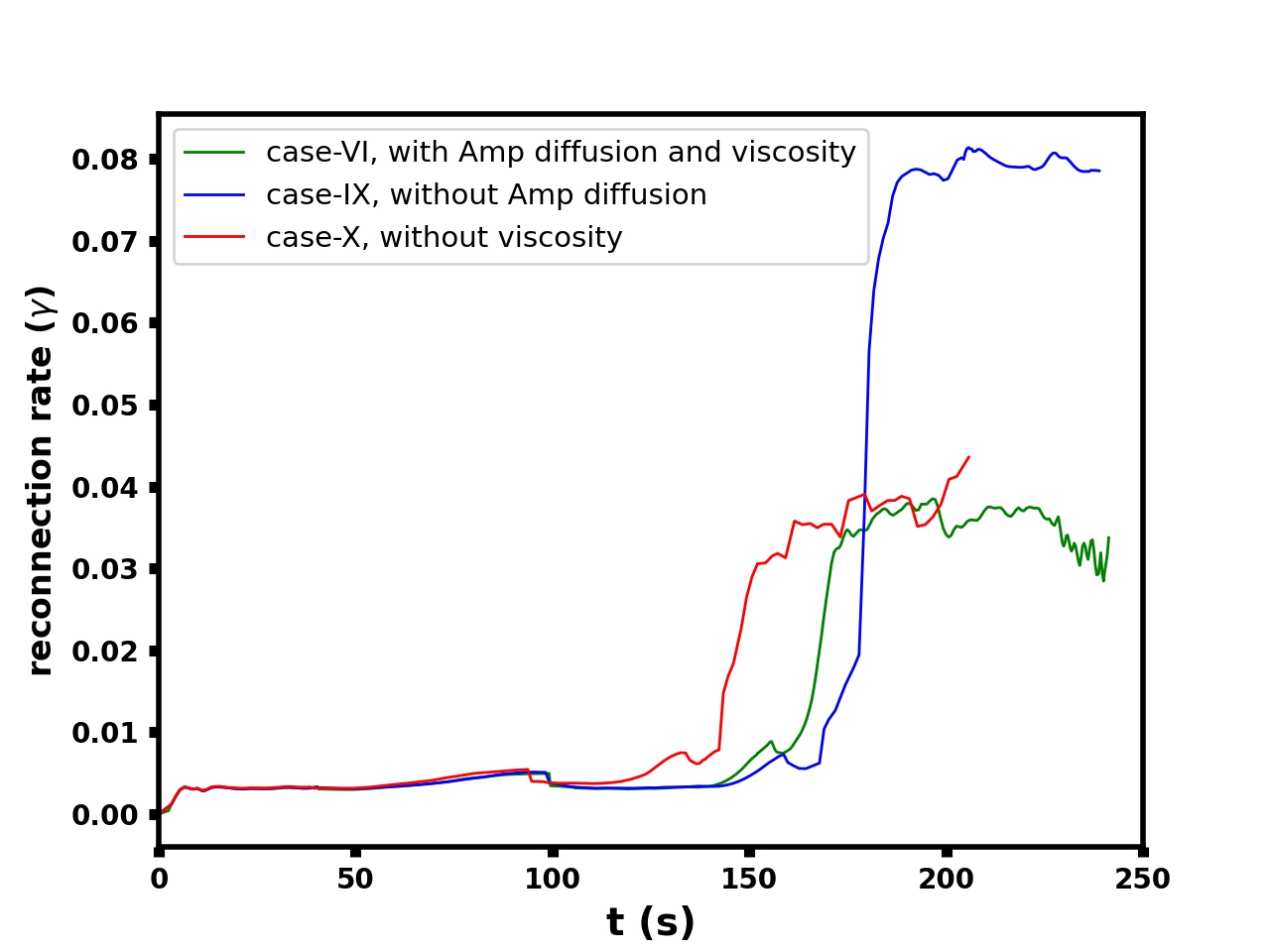}
\put(-130,135){\textbf{(f) Z = 2000 km}}
\end{minipage}
\caption{Temporal evolutions of the reconnection rates in cases with and without radiative cooling models at Z = 400 km (panel a), Z = 800 km (panel c) and Z = 2000 km (panel e), with and without magnetic diffusion contributed by electron neutral collision ($\eta_{en}$) at Z = 400 km (panel b) and Z = 800 km (panel d) and with and without ambipolar diffusion and viscosity at Z = 2000 km (panel f). The effect of radiative cooling and $\eta_{en}$ on reconnection rates decreases with height. With the exclusion of ambipolar diffusion reconnection rate increases significantly for the upper chromospheric case.}
\label{fig_5}
\end{figure}

Comparing the time-dependent reconnection rates at the photospheric height of 400 km above the solar surface for Case I (with radiative cooling and $\eta_{en}$), Case VII (without radiative cooling)  and Case XII (without $\eta_{en}$) in Figs.~\ref{fig_5}(a) and \ref{fig_5}(b), one can find that both the strong radiative cooling and the magnetic diffusion caused by electron-neutral collisions $(\eta_{en})$ have strong effects on magnetic reconnection in the photosphere, they induce the plasmoid instability earlier resulting in significantly higher reconnection rates.
Results shown in Figs.~\ref{fig_5}(a), \ref{fig_5}(c) and \ref{fig_5}(e) indicate that the effect of radiative cooling on reconnection rates becomes weaker when the reconnection region moves to the higher altitude. 
The effect of $\eta_{en}$ on magnetic reconnection also weakens
at higher altitudes (see Figs.~\ref{fig_5}(b) and ~\ref{fig_5}(d)) and can be ignored above the middle chromosphere as shown in Fig.~\ref{fig_4}.

\begin{figure}
\centering
\begin{minipage}{0.49\textwidth}
\includegraphics[width=1.0\textwidth]{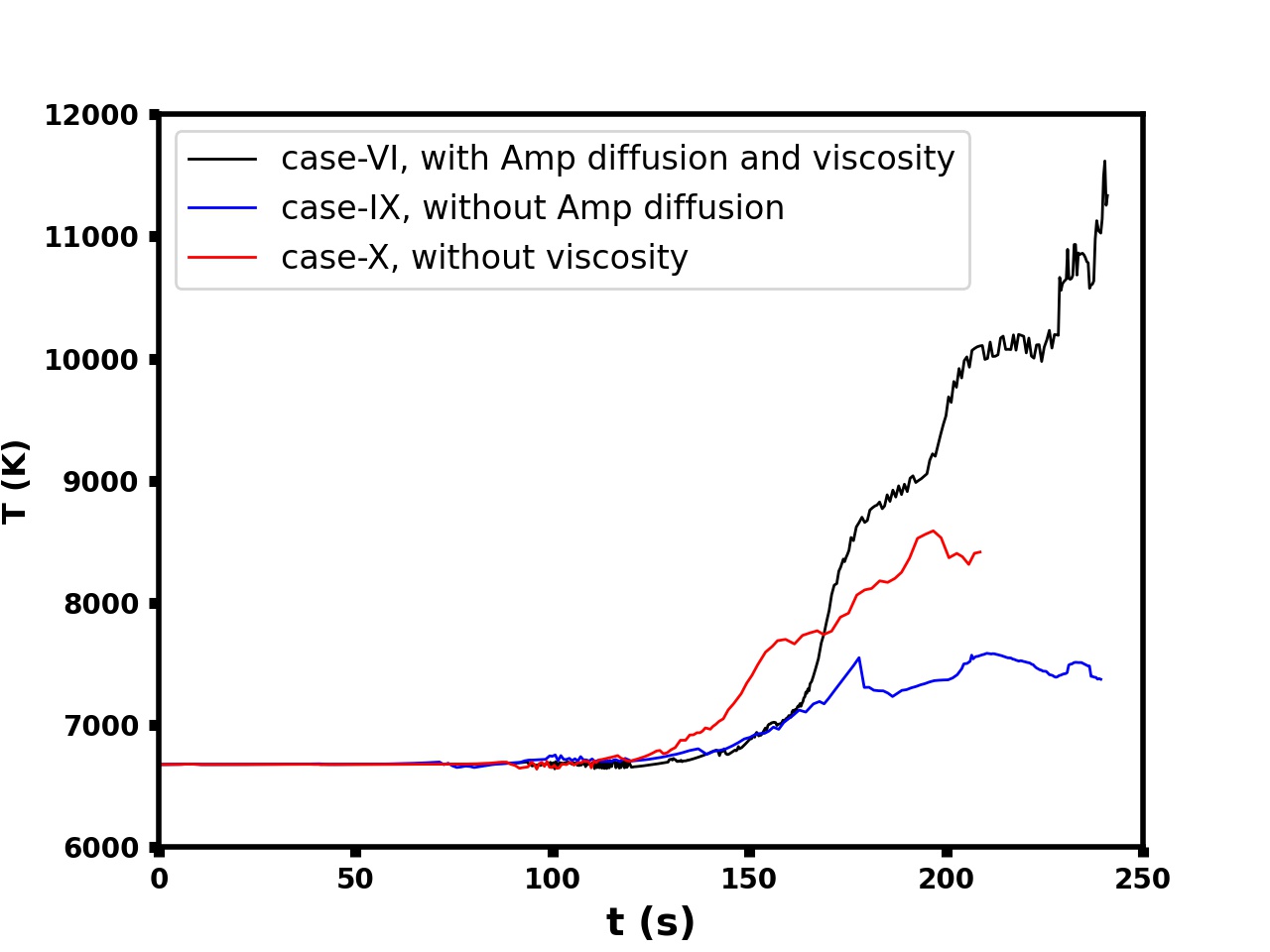}
\put(-110,135){\textbf{(a)}}
\end{minipage}
\begin{minipage}{0.49\textwidth}
\includegraphics[width=1.0\textwidth]{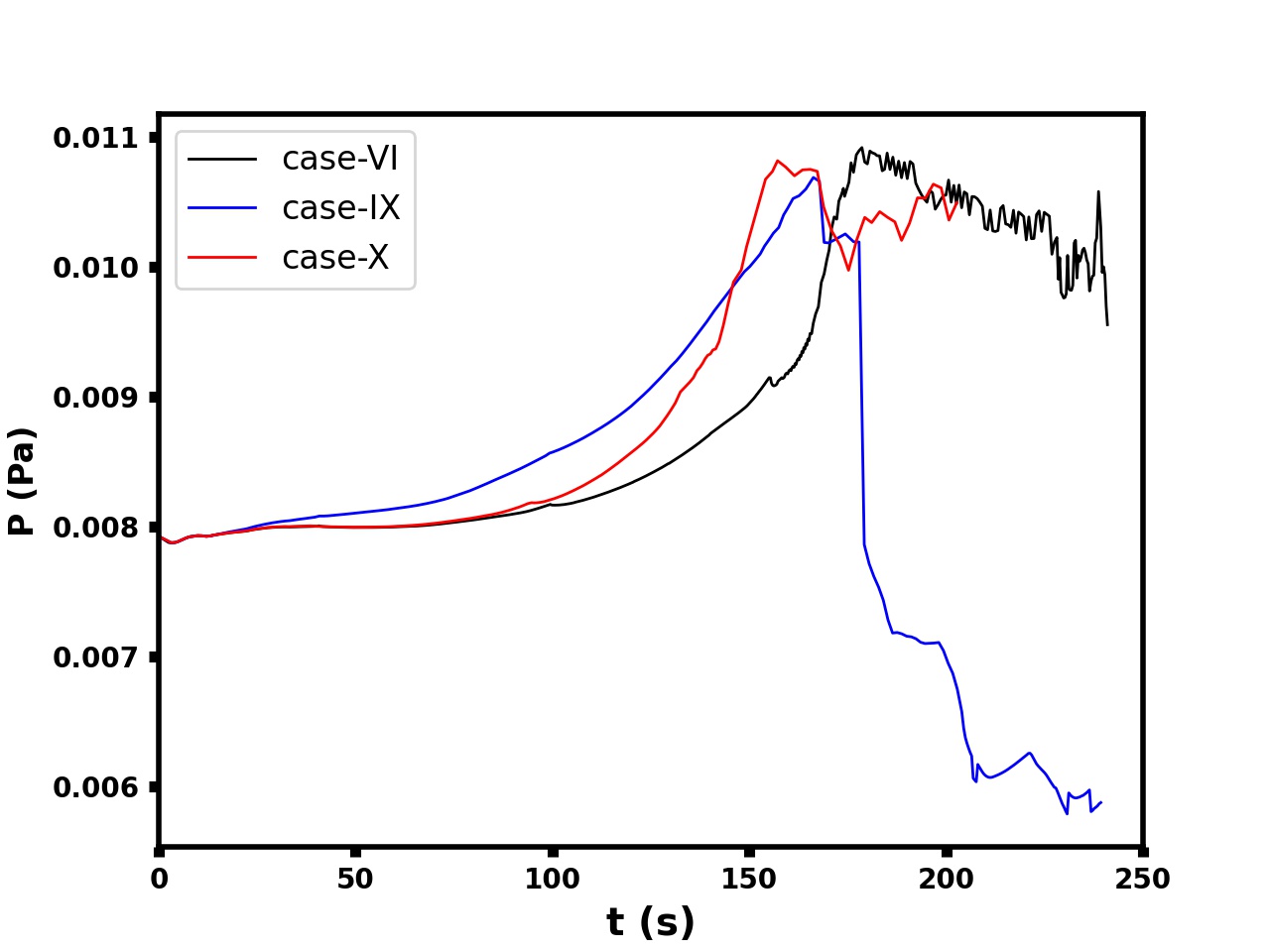}
\put(-110,135){\textbf{(b)}}
\end{minipage}
\caption{Temporal evolutions of the temperature (a) and gas pressure (b) at the main reconnection X-point in cases with and without ambipolar diffusion and viscosity at Z = 2000 km. With ambipolar diffusion and viscosity the temperature and pressure is higher.}
\label{fig_6}
\end{figure}

Figures~\ref{fig_4}(e) and \ref{fig_4}(f) show that the ambipolar diffusion is several orders of magnitude higher than the other diffusivities in the upper chromosphere. 
Therefore, we expect that the ambipolar diffusion might play an important role in magnetic reconnection in this region. 
The existing theoretical results showed that the ambipolar diffusion may cause the formation of a thinner current sheet and higher reconnection rates for the Sweet-Parker type current sheet without a guide field~\cite{brandenburg1994formation}. 
Moreover, the 2D simulations demonstrated that the ambipolar diffusion indeed caused the thinning of the current sheet faster in the case with zero guide field, but the reconnection rate always sharply increases to a high value by the plasmoid instability eventually, the ambipolar diffusion does not cause faster reconnection during such an unstable reconnection period~\cite{ni2015fast}. 
Comparing the reconnection rates at the top of the chromosphere for Case VI (with ambipolar diffusion and viscosity, green line) and Case IX (without  ambipolar diffusion, blue line) in Fig.~\ref{fig_5}(f), we find that including the ambipolar diffusion (Case VI) results in a much smaller reconnection rate after plasmoid instability takes place, the reconnection rate is apparently larger in Case IX without ambipolar diffusion even though the ambipolar diffusion indeed causes the plasmoid instability to occur a little bit earlier in Case VI. 
As shown in Fig.~\ref{fig_6}, including the ambipolar diffusion in Case VI leads to higher temperatures and gas pressures at the principal X-point, making plasma compression in the reconnection region more difficult, slow inflow velocity and low reconnection rate. 
Therefore, we conclude that the ambipolar diffusion effect cannot significantly accelerate the magnetic reconnection process in the low atmosphere, the strong ambipolar diffusion in the upper chromosphere even causes a slower reconnection rate after plasmoid instability in the high $\beta$ reconnection process.

Comparing the reconnection rates in case-VI (green line) and case-X (Z = 2000 km, without viscosity, red line) in Fig.~\ref{fig_5}(f), we find that including the viscosity in Case VI causes the plasmoid instability to occur later. 
The main reason is that the strong viscosity in Case VI leads to a higher temperature and pressure in the reconnection region, which then makes thinning of the current sheet more difficult.

\subsection{Heating mechanisms during magnetic reconnection process at different layers}
Magnetic reconnection is believed to be the main mechanism to heat the transient events in the low solar atmosphere. 
However, we still know very little about the energy conversion mechanisms inside the reconnection diffusion region. 
In this work, we performed high resolution MHD simulations with realistic magnetic diffusions, ambipolar diffusion, viscosity and suitable radiative cooling, which allows us to deeply analyze the heating mechanism of the high $\beta$ reconnection events at different altitudes of the low solar atmosphere.

As seen in Fig.~\ref{fig_3}, the local compression heating inside the current sheet is always the dominant mechanism to heat the plasma after plasmoid instability takes place in all the cases at different altitudes.
Previous high $\beta$ simulations for EBs~\cite{liu2023numerical} and low $\beta$ simulations for UV bursts~\cite{ni2022plausibility} also proved that such a heating mechanism is the dominant one during the later unstable reconnection process. 
The detailed analysis in these previous papers showed that interactions and coalescence of the plasmoids can strongly enhance the local compression inside the reconnection region, the kinetic energy is converted into the thermal energy during this process. 
Our results further confirm that such a mechanism plays an important role in heating events at different altitudes.

Comparing the average power densities during the later stage in different cases, we see that the magnitude of the average power density contributed by Q$_{comp}$, Q$_{ei}$ and Q$_{en}$, as well as the radiative cooling term (Q$_{rad}$), decreases by about six orders of magnitude when we move from the photosphere to the top of the chromosphere. 
The value of $Q_{Amp}$ in the chromosphere is only about two orders of magnitude larger than that in the photosphere. Whereas, the heating contributed by viscosity (Q$_{visc}$) does not change significantly with height and is quantitatively comparable in all simulation runs. 
The amplitudes of Q$_{visc}$ and Q$_{Amp}$ are much smaller than the predominant heating contributor like Q$_{comp}$ in the very low atmosphere (photospheric and lower chromospheric cases). 
However, we cannot ignore the significance of both Q$_{visc}$ and Q$_{Amp}$ in the middle/upper chromosphere layers since their amplitudes are the same as Q$_{comp}$ at higher altitudes. 
Therefore, we conclude that, in addition to compression heating, viscous heating and the ambipolar diffusion heating also play a vital role in heating the lower density plasma in the middle and the upper chromosphere regions.

As shown in Figs.~\ref{fig_3}(a)-\ref{fig_3}(c), we find that the joule heating due to collisions between electrons and neutrals (Q$_{en}$) is dominant over other heating terms in the early stage of magnetic reconnection process at the photosphere and the low chromosphere (Cases I-III). 
This finding is consistent with the recent studies about the heating mechanisms in EBs~\cite{liu2023numerical}, we further prove that the Joule heating contributed by $\eta_{en}$ is very important during the quasi-steady state of reconnection process below the middle chromosphere.

\begin{figure}
\centering
\begin{minipage}{0.329\textwidth}
\includegraphics[width=1.0\textwidth]{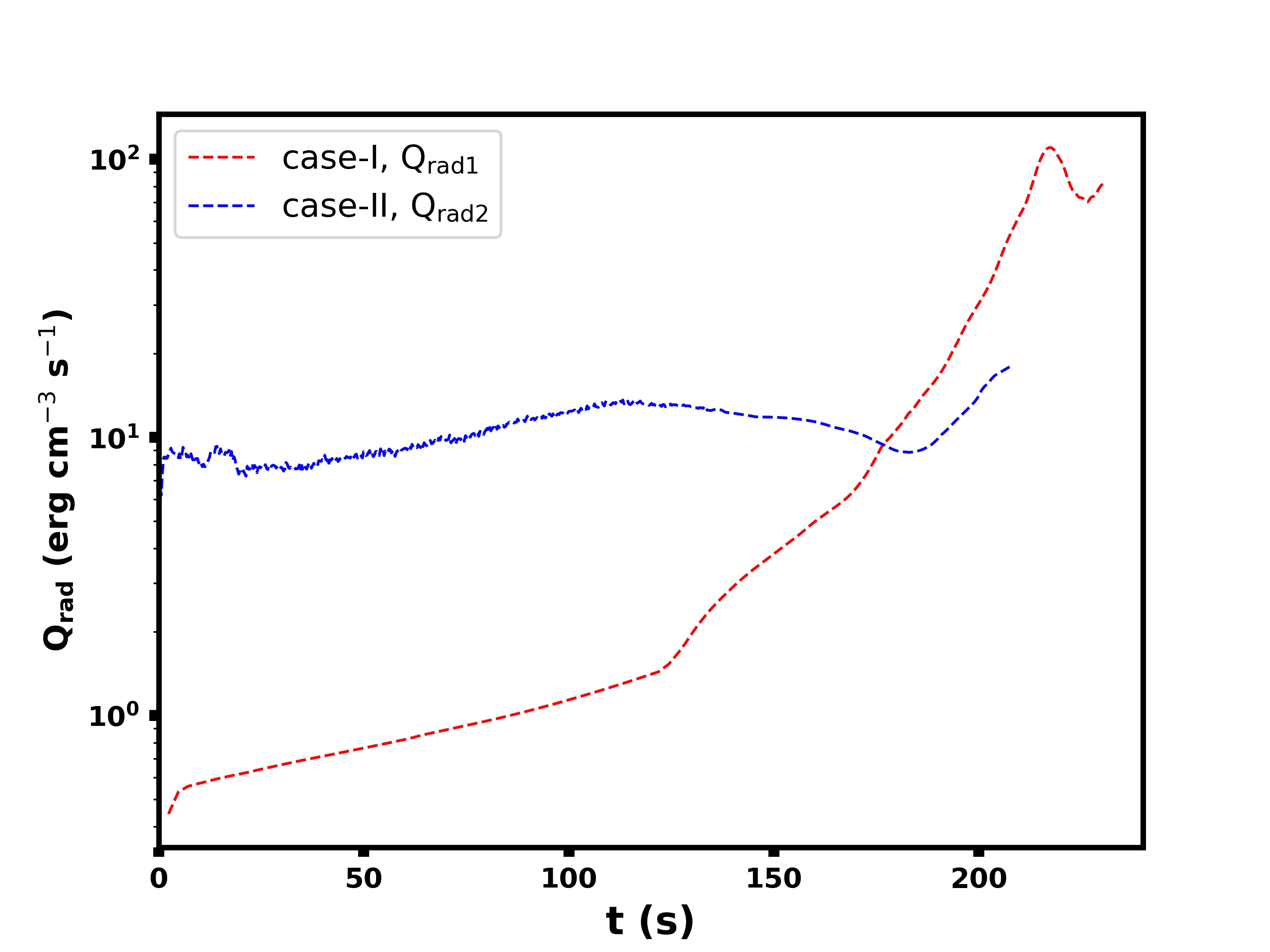}
\put(-95,90){\textbf{(a) Z = 400 km}}
\end{minipage}
\begin{minipage}{0.329\textwidth}
\includegraphics[width=1.0\textwidth]{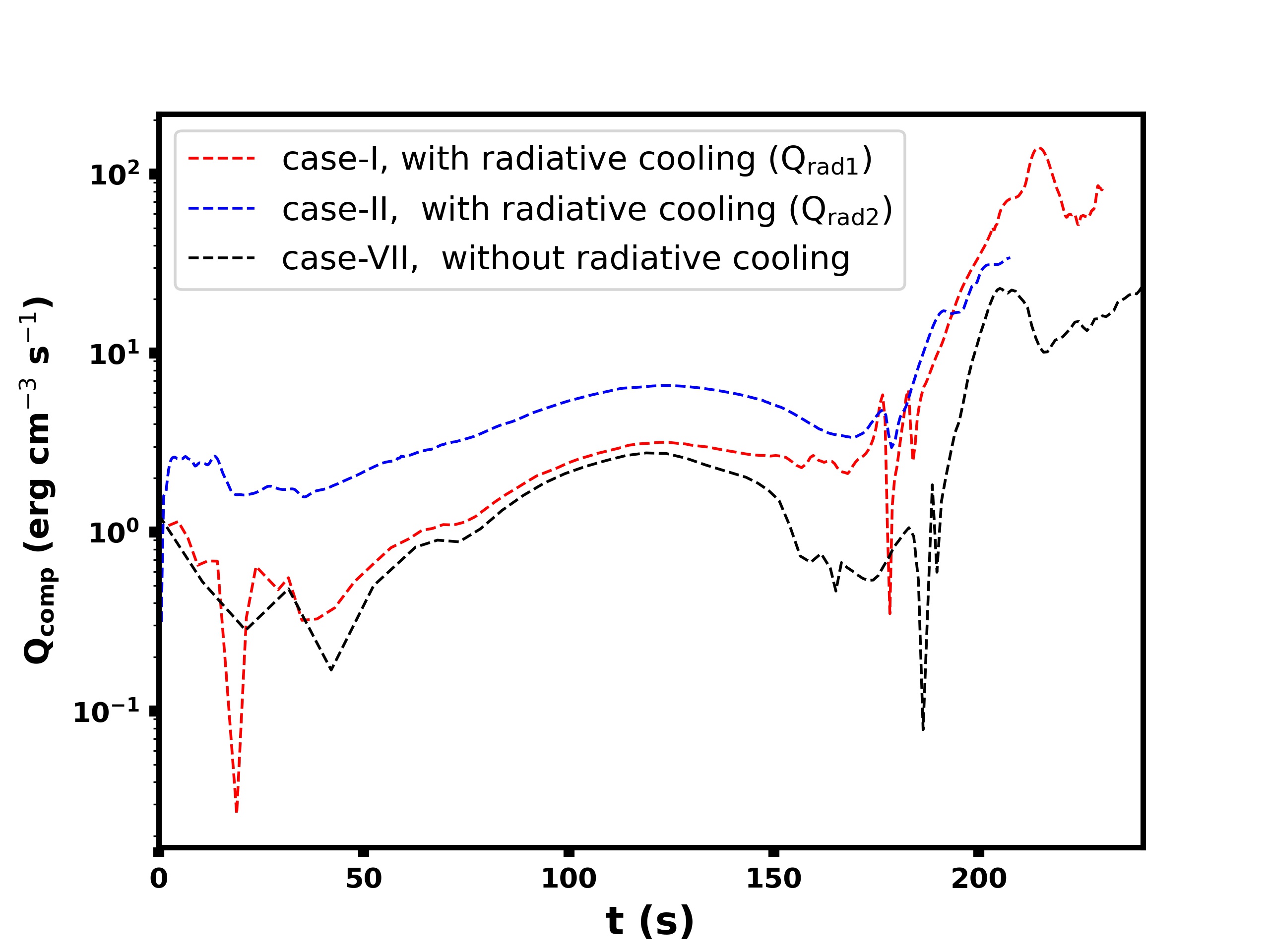}
\put(-95,90){\textbf{(b) Z = 400 km}}
\end{minipage}
\begin{minipage}{0.329\textwidth}
\includegraphics[width=1.0\textwidth]{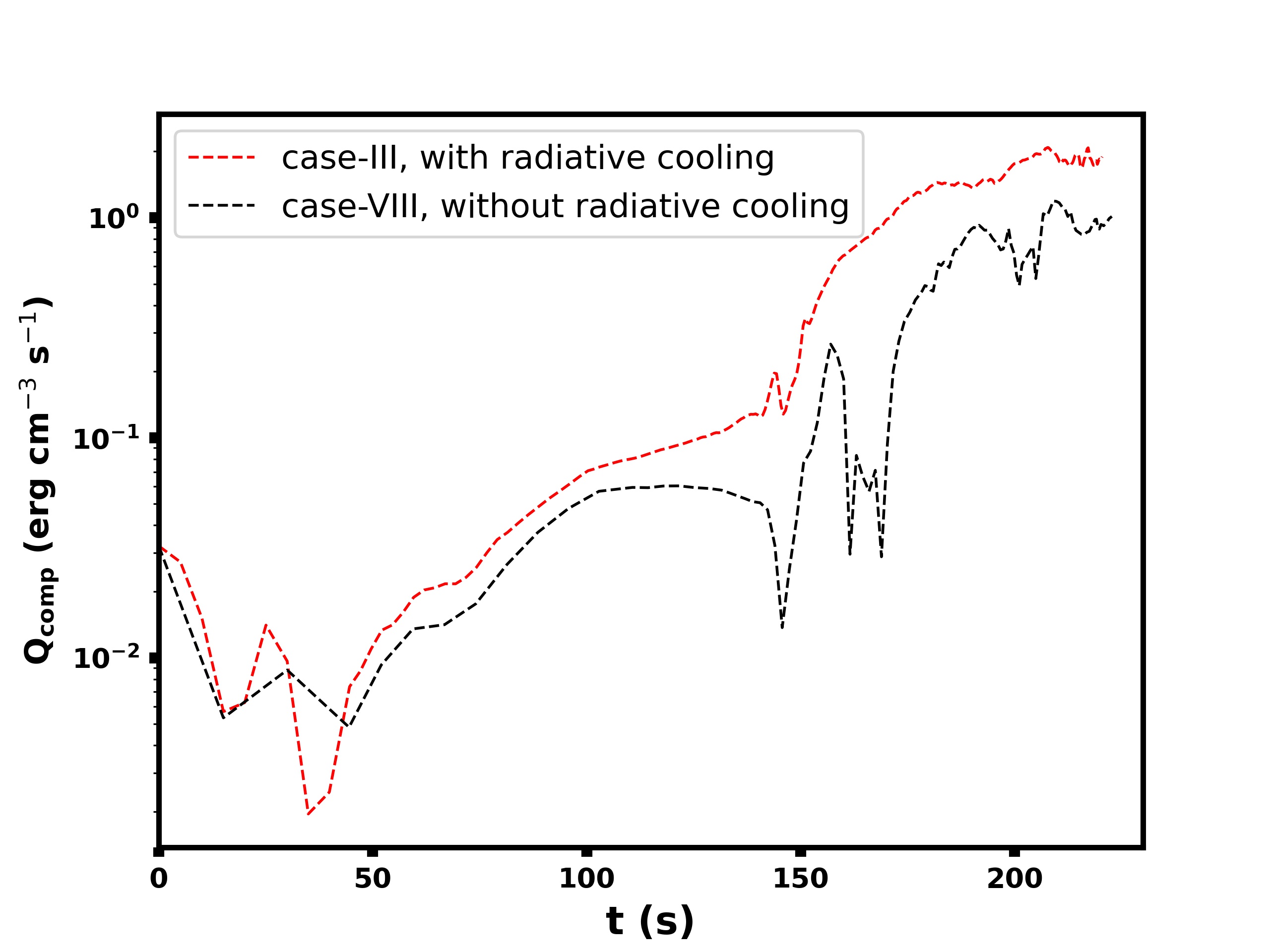}
\put(-95,90){\textbf{(c) Z = 800 km}}
\end{minipage}
\caption{Temporal evolutions of the average power densities contributed by the radiative cooling at Z = 400 km, cases I and II (a), the compression heating at Z = 400 km, cases I, II and VII (b), and the compression heating at Z = 800 km, cases III and VIII (c). Compression heating is stronger in the presence of strong radiative cooling.}
\label{fig_7}
\end{figure}

We also note that the radiative cooling has strong effects on the energy conversion process inside the reconnection region in the low atmosphere. 
The strong radiative cooling results in large values of $\eta_{ei}$ , $\eta_{en}$ and $\eta_{Amp}$, generating more Joule heating and ambipolar diffusion heating. 
As shown in Fig.~\ref{fig_7}, the average power density of the radiative cooling (Q$_{rad}$) in Case I (Z = 400 km, Q$_{rad1}$) is about one order of magnitude smaller than that in Case II (Z = 400 km, Q$_{rad2}$) before the plasmoid instability, which leads to a weaker compression heating (Q$_{comp}$) in Case I during this stage. 
Then, Q$_{rad}$ in photospheric reconnection case with Q$_{rad1}$ (Case I) quickly increases to a higher value after the unstable reconnection process starts, the compression heating Q$_{comp}$ in Case I is also enhanced accordingly. 
The compression heating is the weakest when the radiative cooling is turned off in Case VII (Z = 400 km, without radiative cooling model). 
When the reconnection region moves to higher altitudes, the low chromosphere, we still find that including the radiative cooling effect in Case III (Z = 800 km, Q$_{rad3}$) results in a larger value of Q$_{comp}$ as shown in Fig.~\ref{fig_7}(c). 
These results demonstrate that the stronger the radiation is, the more apparent the compression heating is, the radiation helps generate thermal energy in the reconnection region in the low atmosphere.

\section{Summary}
\label{sec_V}
We have studied magnetic reconnection at different altitudes in the cool low solar atmosphere while considering time-dependent ionization of the partially ionized hydrogen-helium plasma. 
The simulations were carried out in high-$\beta$ plasma with weak magnetic field using single-fluid MHD model, and suitable radiation models are included. 
The study explores the significance of diffusivities, viscosity and radiation losses on magnetic reconnection at different altitudes. The main conclusions are as follows:
\begin{enumerate}
    \item Plasmoid instability plays a vital role in leading to fast magnetic reconnection at different altitudes. However, both the strong radiative cooling and magnetic diffusion caused by the electron-neutral collisions significantly accelerate the magnetic reconnection process below the middle chromosphere. Both ambipolar diffusion and viscosity result in higher temperature and plasma pressure in the reconnection region in the upper chromosphere, which then decreases the inflow velocity in the inflow region. Therefore, the strong ambipolar diffusion in the upper chromosphere even suppresses the reconnection process after the plasmoid instability takes place. The viscosity in the upper chromosphere also makes the plasmoid instability occur later.
    \item Interactions and coalescence of plasmoids strongly enhance the local compression inside the reconnection region, which becomes the dominant mechanism to heat the plasma in the unstable reconnection process at different altitudes. Both the viscous heating and the ambipolar diffusion heating play an equally important role as the compression heating in the upper chromosphere. Below the middle chromosphere, the Joule heating contributed by $\eta_{en}$ (magnetic diffusion caused by electron-neutral collisions) dominates the heating of the plasma during the early quasi-steady reconnection stage, and the strong radiative cooling leads to stronger compression heating and more generation of the thermal energy in the whole reconnection process. 
    \item Though the plasma $\beta$ is the same in all the cases at different altitudes, the temperature increase is more significant in the reconnection region with lower plasma density and weaker radiative cooling in the upper chromosphere.
\end{enumerate}
 
In this work, single-fluid MHD simulations of high resolution revealed the fast reconnection mechanism and the heating mechanism in the high $\beta$ environment. 
However, we should also note that the effects of non-equilibrium ionization-recombination might play an important role in magnetic reconnection when the length scale reduces to the mean free-path of the ion-neutral collision, which needs multi-fluid MHD studies in the future. 
When the reconnection region moves to the upper chromosphere, the collisions among different particles become much weaker, the Hall effect might become important for invoking fast magnetic reconnection~\cite{jara2019kinetic}, but the related studies are outside the scope of this work. 
The change in plasma parameters and strength of magnetic fields in different regions causes the plasma $\beta$ to change by several orders of magnitude. 
We plan to assess the effect of low initial plasma $\beta$ on magnetic reconnection and heating mechanisms in the low solar atmosphere in the future studies.

\begin{acknowledgments}
This research is supported by the National Key R\&D Program of China Nos. 2022YFF0503800 (2022YFF0503804) and 2022YFF0503003 (2022YFF0503000); the NSFC Grants 11973083 and 11933009; the Strategic Priority Research Program of CAS with grant XDA17040507; the outstanding member of the Youth Innovation Promotion Association CAS (No. Y2021024); the Applied Basic Research of Yunnan Province in China Grant 2018FB009; the Yunling Talent Project for the Youth; the project of the Group for Innovation of Yunnan Province grant 2018HC023; the Yunling Scholar Project of the Yunnan Province and the Yunnan Province Scientist Workshop of Solar Physics; Yunnan Key Laboratory of Solar Physics and Space Exploration (No. 202205AG070009); The numerical calculations and the data analyses have been done on Hefei advanced computing center and on the Computational Solar Physics Laboratory of Yunnan Observatories.
\end{acknowledgments}

\section*{References}
\nocite{*}
\bibliography{aipsamp}
\end{document}